\documentclass[fleqn,10pt]{SelfArx} 

\usepackage[english]{babel} 
\usepackage{amsmath, amsthm, amssymb, amsfonts}
\usepackage{graphicx}
\usepackage{url}
\usepackage{float}
\usepackage{algpseudocode}
\usepackage{algorithm}

\definecolor{color1}{RGB}{0,0,90} 
\definecolor{color2}{RGB}{0,20,20} 

\usepackage{hyperref} 
\hypersetup{hidelinks,colorlinks,breaklinks=true,urlcolor=blue,citecolor=color1,linkcolor=color1,bookmarksopen=false,pdftitle={Title},pdfauthor={Author}}

\JournalInfo{The Rise and Fall of Network Stars} 
\Archive{Draft Article} 

\PaperTitle{The Rise and Fall of Network Stars: Analyzing 2.5 Million Graphs to Reveal How High-Degree Vertices Emerge over Time} 

\Authors{Michael Fire\textsuperscript{1,2}* and Carlos Guestrin\textsuperscript{1}} 
\affiliation{\textsuperscript{1}\textit{Paul G. Allen School of Computer Science \& Engineering, University of Washington, Seattle, WA}} 
\affiliation{\textsuperscript{2}\textit{The eScience Institute, University of Washington, Seattle, WA}} 
\affiliation{*\textbf{Corresponding author}: fire@cs.washington.edu} 


\newcommand{\keywordname}{Keywords} 

\Abstract{Trends change rapidly in today's world, prompting this key question: What is the mechanism behind the emergence of new trends? By representing real-world dynamic systems as complex networks, the emergence of new trends can be symbolized by vertices that ``shine.'' That is, at a specific time interval in a network's life, certain vertices become increasingly connected to other vertices. This process creates new high-degree vertices, i.e., network stars. Thus, to study trends, we must look at how networks evolve over time and determine how the stars behave. In our research, we constructed the largest publicly available network evolution dataset to date, which contains 38,000 real-world networks and 2.5 million graphs. Then, we performed the first precise wide-scale analysis of the evolution of networks with various scales. Three primary observations resulted: (a) links are most prevalent among vertices that join a network at a similar time; (b) the rate that new vertices join a network is a central factor in molding a network's topology; and (c) the emergence of network stars (high-degree vertices) is correlated with fast-growing networks. We applied our learnings to develop a flexible network-generation model based on large-scale, real-world data. This model gives a better understanding of how stars rise and fall within networks, and is applicable to dynamic systems both in nature and society.
\\ \\
\textbf{Multimedia Links} \\
$\blacktriangleright$ \href{https://youtu.be/zZNZ03ULASk}{\underline{Video}} \hspace{2mm}
\smallskip $\blacktriangleright$ \href{https://dynamics.cs.washington.edu/}{\underline{Interactive Data Visualization}} \hspace{2mm}
 $\blacktriangleright$ \href{https://dynamics.cs.washington.edu/data.html}{\underline{Data}} \hspace{2mm}
 $\blacktriangleright$ \href{https://dynamics.cs.washington.edu/code.html}{\underline{Code Tutorials}}
}

\begin{document}

\flushbottom 

\maketitle 

\tableofcontents 

\thispagestyle{empty} 


\section{\label{sec:intro}Introduction}
Change is inevitable, yet the mechanisms behind changing trends are not well understood ~\cite{barabasi2009scale,prvzulj2016network}. By investigating these mechanisms, we can better answer questions such as how people gain and lose political power, why some companies thrive while others shrivel, and how infectious diseases patterns can spread throughout populations.
To study the mechanisms that influence new trends, we can represent various dynamic systems as complex networks and then explore how these networks change over time. Complex networks are loosely defined as networks with non-trivial structure and dynamics, appearing in many real-world systems~\cite{barabasi2003linked,ravasz2003hierarchical,strogatz2001exploring}. Networks consist of a set of vertices and a set of links connecting these vertices. Vertices can represent a wide range of entities, such as online social network users~\cite{fire2013computationally}, neurons~\cite{bullmore2009complex}, or proteins~\cite{barabasi2009scale,albert2002statistical}. The popularity of a vertex can be measured by the number of links connected to it from other vertices in the network, where the links can be directed, like in Twitter\footnote{\url{http://twitter.com}} where one user follows another user, or undirected, like a mutual friendship between two people~\cite{fire2013computationally}. The most popular vertices--vertices with many connections--are referred to as stars. 
The objective of our research was to use large-scale, real-world data to better understand how real-world networks evolve over long periods of time. We then narrowed that objective to study network stars and gain significant insights into their rise and fall over months, years, and even centuries as networks evolve.

We utilized a variety of large-scale datasets, data science tools, and extensive cloud computing resources to assemble the world's largest complex network
 evolution dataset. The dataset consists of billions of records used to construct and analyze the evolution process of over 38,000 complex networks and the topological properties of more than 2.5 million graphs over long periods of times 
(see Table~\ref{tab:datasets} and Section~\ref{sec:datasets}). Namely, we constructed and analyzed the following networks:
\begin{itemize}
 \item \textit{Citation and co-authorship networks}, created from the Microsoft Academic Graph~\cite{sinha2015overview}, which includes more than 126 million papers and 114 million authors over a period of 215 years. 
 \item The Reddit social network, created from over 2.71 billion comments over a period of more than 10 years~\cite{reddit}. 
 \item \textit{Chess players network}, created from over 214 million games during a period of 18 years~\cite{fics}. 
 \item \textit{People marriage network}, created from the WikiTree online genealogy dataset, including 1.96 million marriage records over 610 years~\cite{wikitree}. 
 \item \textit{Bitcoin network}, created by over 37 million Bitcoin transactions over a period of 4 years~\cite{bitcoin}. 
\end{itemize}
In addition to analyzing the large complex networks described above, we analyzed the evolution of about 18,000 co-authorship and citation networks of
 various research fields. We also analyzed the evolution of more than 20,000 communities for the Reddit dataset (see Section~\ref{sec:reddit}).

\begin{table*}
  \centering
    \caption{Network Datasets.}
  \includegraphics[scale=0.8]{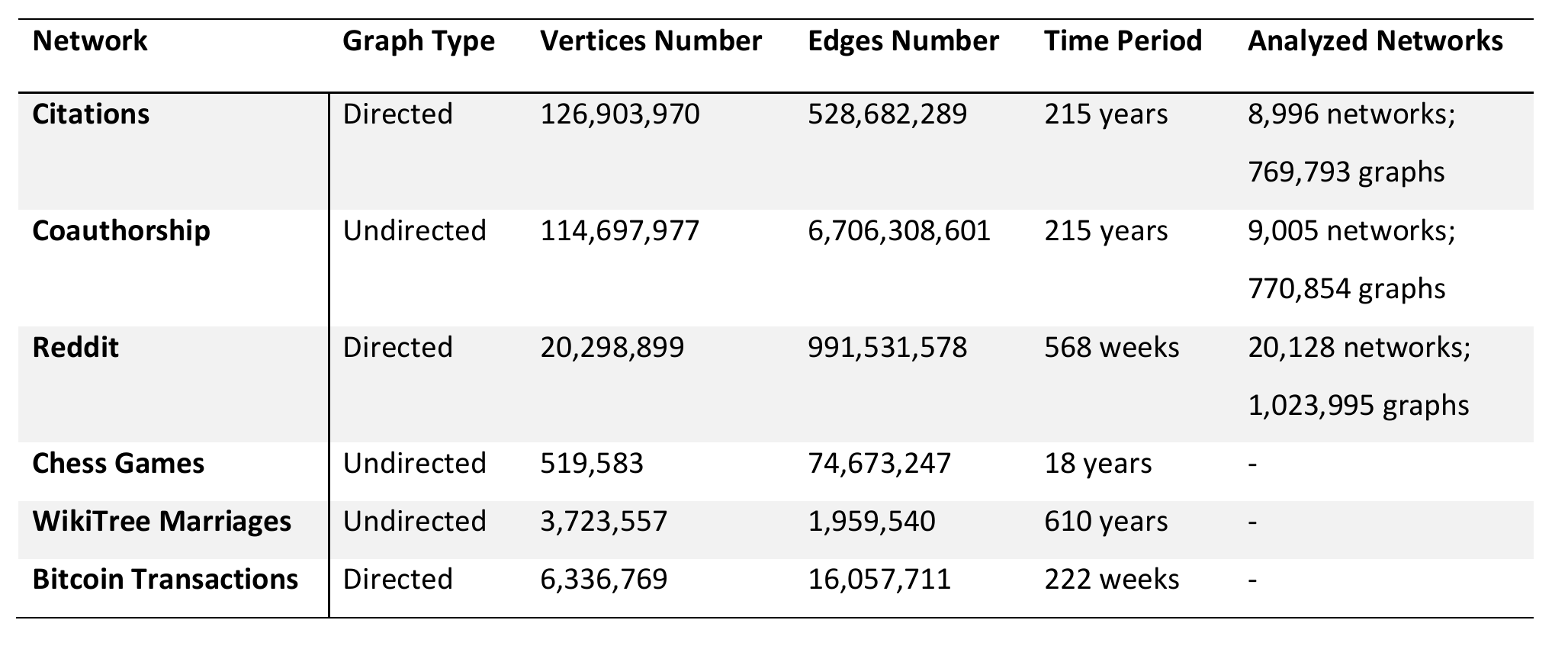}

  \label{tab:datasets}
\end{table*}

We utilized the constructed extensive dataset to perform the first precise wide-scale analysis of the evolution of networks with various scales.
By examining the evolution and dynamics of these networks, three notable observations emerged:
First, links are most prevalent among vertices that join a network at a similar time (see Figure~\ref{fig:time_diff}). 
For example, in the citation network, over 80\% of all citations referenced publications published within 15 years,
 while less than 8\% had a publication gap of more than 25 years. Similarly, in the WikiTree network, 69.2\% and 8.2\% of married couples had age differences of fewer than 7 years or over 15 years, respectively.
 
Second, the rate that new vertices join a network is a central factor in molding a network's topology.
We identified six common patterns in which vertices tend to join the networks (see Figure~\ref{fig:jrcs} ). 
Moreover, we observed that different vertex-join patterns influence the structures and properties of the networks (see Figures~\ref{fig:corr} and~\ref{fig:fast_slow}). 
For example, we identified that on average fast-growing networks tend be active longer, have more vertices, be less dense, and cluster less than slow-growing networks.

Third, network stars (high-degree vertices) tend to emerge in networks that are growing rapidly. 
For slow-growing networks, most stars emerged a short time after the network became active and kept their place, while for fast-growing networks, stars emerged at any time (see Figure~\ref{fig:stars_heatmap}).

We applied our learnings to develop a straightforward random network-generation model that more accurately depicts how networks evolve (see Figures~\ref{fig:tpa} and~\ref{fig:tpa_example}, and Table~\ref{tab:tpa}).
 Our Temporal Preferential Attachment (TPA) model improves upon previous models because it more correctly represents real-world data,  especially for networks that are growing quickly, and can be used in a more flexible manner. Furthermore, our model can give insights on the changing popularity of network stars.

This study has several contributions. To our knowledge, this is the largest study--by several orders of magnitude--to analyze real-world complex networks over long periods of times.

The key contributions presented in this paper are fivefold:
First, we constructed the largest network evolution corpora that is publicly available. The dataset consists of billions of records that we used to construct and analyze the evolution process of over 38,000 complex networks and the topological properties of more than 2.5 million graphs over long periods of times.  This dataset can immensely aid researchers in investigating and understanding complex dynamic systems.

 Second, we observed that time is a crucial factor in the way a network evolves. Vertices tend to connect to other vertices that join the network at a similar time. For example, in the citation network, over 80\% of all citations referenced publications published within 15 years,
 while less than 8\% had a publication gap of more than 25 years.

 Third, we found that the rate new vertices join a network is a central factor in molding a network's topology.
We identified six common patterns in which vertices tend to join a network (see Figure~\ref{fig:jrcs} ). 
Moreover, we observed that different vertex-join patterns influence the structure and properties of a network (see Figures~\ref{fig:corr} and~\ref{fig:fast_slow}). 
For example, we identified that on average fast-growing networks tend be active longer, have more vertices, be less dense, and cluster less than slow-growing networks. 

Fourth, we discovered that network stars (high-degree vertices) tend to emerge in networks that are growing rapidly. 
For slow-growing networks, most stars emerged a short time after the network became active and kept their place, while for fast-growing networks, stars emerged at any time (see Figure~\ref{fig:stars_heatmap}).

Fifth, we developed a simple model, utilizing all the above observations, that uses real-world big data to confirm and explain our observations. Our Temporal Preferential Attachment (TPA) model improves upon previous models because it more correctly represents real-world data, especially for networks that are growing quickly, and can be used in a more flexible manner. Furthermore, our model gives insights on the changing popularity of network stars. 
  
The remainder of the paper is organized as follows: In Section~\ref{sec:related}, we provide an overview of various related studies. 
In Section~\ref{sec:methods}, we describe the datasets, methods, algorithms, and experiments used throughout this study. 
Next, in Section~\ref{sec:results}, we present the results of our study.
Afterwards, in Section~\ref{sec:tpa}, we present our TPA model.
 Then, in Section~\ref{sec:diss}, we discuss the obtained results. 
 Lastly, in Section~\ref{sec:con}, we present our conclusions from this study and also offer future research directions.

\section{Related Work}
\label{sec:related}
The study of complex networks began over half a century ago, in 1965. 
While studying a network of citations among scientific papers, Price observed a network in which the degree distribution followed a power law~\cite{price1965statistical}.
 Later, in 1976, Price~\cite{price1976general} provided an explanation of the creation of these types of networks: ``Success seems to breed success. 
 A paper which has been cited many times is more likely to be cited again than one which has been little cited''~\cite{price1976general}. 
 Price subsequently offered a method for the creation of networks in which the degree distribution follows a power law.

Several decades later, Watts and Strogatz~\cite{watts1998collective} and Newman and Watts ~\cite{newman1999renormalization} introduced models for generating small-world networks. 
Typically, small-world networks have a relatively high clustering coefficient, and the distance between any two vertices scales as the logarithm of the number of vertices~\cite{sallaberry2013model}. 
Barab{\'a}si and Albert observed that degree distributions that follow power laws exist in a variety of networks,
 including the World Wide Web~\cite{barabasi1999emergence}. Barab{\'a}si and Albert coined the term ``scale-free networks'' for describing such networks.
Similar to Price's method~\cite{price1976general}, Barab{\'a}si and Albert~\cite{barabasi1999emergence} suggested a simple and elegant model
 for creating random complex networks based on the rule that the rich are getting richer. 
 In the BA model, a network starts with $m$ connected vertices. Each new vertex that is added (one at a time) has a greater probability of connecting to
  pre-existing vertices with higher degree, where the probability of connecting to an existing vertex is proportional to vertex's degree~\cite{barabasi1999emergence}. 
  Consequently, rich vertices with high degrees tend to become even richer due to their connections with new vertices that join the graph. 
  Many real-world complex networks have a community structure in which ``the division of network nodes into groups within which the network connections are dense, but between which they are sparser''~\cite{newman2004finding}. 
In 2000, Dorogovtsev et al.~\cite{dorogovtsev2000structure} suggested a model with preferential 
linking that takes into consideration  a vertex attractiveness. 
In 2002, Holme and Kim~\cite{holme2002growing} extended  the Barab{\'a}si and Albert model to include a ``triad formation step.'' The Holme and Kim model creates networks with both the perfect power-law degree distribution and high clustering. In 2004, Newman and Girvan proposed a community detection algorithm and offered a simple method to create networks with community structure~\cite{newman2004finding}. In 2007, Leskovec et al.~\cite{leskovec2007graph} introduced the ``forest fire'' graph generation model, based on a ``forest fire'' spreading process.
 
Even though the models described above can explain some of the characteristics of real-world complex networks, 
the random networks created by these models were lacking in other properties that were observed in real-world complex networks. 
Therefore, in recent years, other models have been suggested which have additional 
characteristics~\cite{watts1998collective,sallaberry2013model,holme2002growing}.
Thorough reviews on complex networks and complex network evolution models can be found in books by Chung and Lu~\cite{chung2006complex}, 
Newman et al.~\cite{newman2011structure},  and by Dorogovtsev and Mendes~\cite{dorogovtsev2013evolution}.

A similar study to ours was conducted by Leskovec et al.~\cite{leskovec2008microscopic}. 
They performed edge-by-edge analysis of four large-scale networks – Flickr, Delicious, LinkedIn, and Yahoo Answers – 
with time spans ranging from four months to almost four years. 
By studying a wide variety of network formation strategies, they observed that edge locality
plays a critical role in the evolution of networks, and they offered a model which focused on microscopic vertex behavior. 
In their proposed model, vertices arrive at a pre-specified rate and choose their lifetimes. Afterwards, each vertex ``independently initiates edges according to a `gap'
 process, selecting a destination for each edge according to a simple triangle-closing model free of any parameters''~\cite{leskovec2008microscopic}. 
 They showed that their model could closely mimic the macroscopic characteristics of real social networks.
Additionally, Leskovec et al., similar to our study, observed the arrival patterns of various vertices.
 Namely, they observed that (a) Flickr's network data has grown exponentially; (b) Delicious has grown slightly superlinearly; (c) LinkedIn has
grown quadratically; and (d) Yahoo Answers has grown sublinearly. Due to these observations, they concluded that vertex arrival functions needed to be part of their proposed model. 
However, their study did not analyze the implications of using different arrival functions.

The body of literature has increased extensively over the last two decades, with hundreds of new network studies each year,\footnote{According to Google Scholar over 870 papers' titles 
published in 2016 contains the phrase ``complex networks.''} and many 
papers present observations and network models that overlap with this study. 
To the best of our knowledge, however, this study is the first to present a general model 
based on extensive analysis of large-scale real data utilizing over 38,000 real-world complex networks. Moreover, this study is the first to utilize extensive
temporal complex network data to understand how high-degree vertices emerge over time.

\section{Methods and Experiments}
\label{sec:methods}
\subsection{Constructing the Network Datasets}
\label{sec:datasets}
In this study, we utilized six different datasets to construct various types of networks. 
Below we describe in detail how we generated the complex network corpora with over 38,000 networks.

\subsubsection{The Reddit Networks}
\label{sec:reddit}
Reddit is a news aggregation website and online social platform launched in 2005 by Steve Huffman and Alexis Ohanian~\cite{bergstrom2011don}. 
Reddit users (also known as ``redditors'') can submit content on the website, which is then commented upon,
 and upvoted or downvoted by other users in order to increase or decrease the submission visibility.
 Redditors can also create their own subreddit on a topic of their choosing, make it public or private, and let other redditors join it. 
This makes Reddit a collection of online communities centered around a variety of topics such as books, gaming, science, and asking questions
In this study, we utilized the Reddit dataset which was recently made public by Jason Michael Baumgartner~\cite{reddit}.
Specifically, we utilized over 2.71 billion comments that were posted from December 2005 through October 2016.
These posts were created by 20,299,812 users with unique usernames in 416,729 different subreddits.
The dataset contains information on the exact time and date each comment was posted.
Moreover, the dataset contains each comment's ID, as well as information on the user who posted it and the ID of the parent comment,
i.e., the ID to which the current comment replied.
We cleaned the dataset by removing nonessential comments, specifically those that were marked as deleted and those that did not include the information
of the user who posted them.
Additionally, we removed posts by users who with high probability were bots.
Namely, we removed all the users who posted more than 100,000 comments each, 
and we removed redditors whose comments appeared in the bots list published in the BotWatchman 
subreddit.\footnote{\url{https://www.reddit.com/r/BotWatchman/}}
We downloaded the bots list from the BotWatchman subreddit during November 
2016.
After the removal of these posts, we were left with over 2.39 billion comments published in 371,841 subreddits by 20,298,899 users.

Next, we constructed social networks from the subreddits' comments data.
However, many of the subreddits did not contain enough comments. 
Therefore, for all the subreddits in the clean dataset with about 2.39 billion comments,
we selected only those subreddits that had at least 1,000 comments and more than a single user.
Out of all the subreddits, 20,145 fulfilled these criteria, 
out of which we succeeded in constructing the social networks over time of 20,136 subreddits with over 2.37 billion posts 
(referred to as selected subreddits). Afterwards, for each selected subreddit, similar to the construction method used by Kairam et al.~\cite{kairam2012life},
 we created the subreddit's social network directed graph by connecting users who posted comments as replies to other posted comments. 
 
Namely, for a subreddit $S$, we define the subreddit's directed graph at time $t$ to be: 
$G_t^S:=<V_t^S,E_t^S>$, where $V_t^S$
is the set of vertices representing all the subreddit's users who posted at least a single comment in the subreddit up to $t$ 
days after the subreddit became active, i.e., when the first comment was published in the subreddit $S$.
In addition, $e:= (u,v) \in E_t^S$ is the list of all edges between the subreddit's users, $u \in V_t^S$  and $v \in V_t^S$,
 created up to $t$ days after the subreddit became active. We define an edge between $u$ and $v$ to exist if there exists a comment on the subreddit posted by $u$ 
 to which $v$ posted a reply on the same subreddit. Lastly, to better understand how subreddits evolve over time, for each selected subreddit $S$, 
 we created a set of incremental graphs in incremental time intervals of every 4 weeks between the time the subreddit 
 initially became active and the time the last comment was posted in the subreddit according to the dataset. 
 Overall, we created over a million graphs that contain detailed information on how these selected subreddits evolved over time.
 
It is important to notice that the constructed directed graphs also include single
vertices of redditors who posted comments and did not receive any reply, as well as
self-loop edges of redditors who posted a comment and then posted a reply to their own comment.

\subsubsection{The Free Internet Chess Server Network}
\label{sec:chess}
The Free Internet Chess Server (FICS)\footnote{\url{www.freechess.org/}} is one of the oldest and largest Internet chess servers. 
The FICS serves over 540,000 users who have played over 300 million chess games~\cite{fics}. 
For this study, we downloaded the details of 214,873,738 chess games played between January 1999 and January 2016 from the FICS Games Database website~\cite{fics}.
We then extracted each game's metadata, which included the users who played the game and the time the game was played.
Using these details, we constructed a complex network $G_t^C: = <V_t^C,E_t^C>$, in which the vertices $V_t^c$ is a set of all FICS users in our dataset, 
 and $e := (u,v) \in E_t^C$ is the list of all edges between the FICS users $u \in V_t^c$  and $v \in V_t^c$, where 
$u$ and $v$ played at least one game on FICS during the $t$ weeks since the first game in our dataset.
  To study how the chess games network evolves over time, we constructed the network's graph every 4 weeks over a period of 18 years. 

\subsubsection{The WikiTree Marriage Network}
\label{sec:wikitree}
We constructed a large social network using online genealogical records obtained from the WikiTree website~\cite{wikitree}. 
WikiTree is an online genealogical website, created by Chris Whitten in 2008, with a mission to create a single worldwide family tree 
that will make genealogy free and accessible. The website contains over 13 million profile pages of people who lived in the previous centuries, 
and many of the profiles contain specific details about each individual, 
including full name, gender, date of birth, children's profiles, and spouses' profiles. 
To keep WikiTree's data integrity, only invited users can contribute, and contributors must agree to follow an honor code which specifies
 how they should treat openness, accuracy, mistakes, and giving credit.  
 Moreover, many profiles reference the source of the data presented in the profile. 
 Additionally, most profiles have a manager who has responsibility for WikiTree profiles~\cite{WikiTreeManager}, 
 and each profile has its own ``Trusted List'' of people who have access to modify the profile,
  making the information in many profiles only editable to a limited number of people~\cite{WikiTreeTrust}.

In 2015, Fire and Elovici~\cite{fire2015data} showed that it possible to utilize WikiTree data to create a large-scale social network 
that can be used to better understand lifespan patterns in human population. 
Similar to Fire and Elovici's study, in this study we utilized WikiTree data, which was downloaded in April 2016
and includes 1,964,331 marriage records of people whose birth years were between 1400 and 2010, 
to construct the marriage social network. Namely, we constructed a WikiTree marriage network graph at year y to be:
$G_y^W := <V_y^W,E_y^W>$, where $V_y^W$ is a set of people who, 
according to WikiTree's records, were born after 1400 and were married at least once before or during the year $y$, and each link, $e := (u,v)\in E_y^W$, 
is between two individuals, $u,v \in V_y^W$, who got married before or during the year $y$. 

\subsubsection{The Co-authorship Networks}
\label{sec:coauthors}
The Microsoft Academic Graph is a large-scale dataset which contains scientific publication records of 126 million papers, 
along with citation relationships between those publications, as well as relationships between authors, institutions, journals, conferences, and fields of study~\cite{sinha2015overview}. 
The dataset also contains field-of-study hierarchy with four levels, L0 to L3, where L0 is the highest level, such as a research field of Computer Science, 
and L3 is the lowest level, such as a research field of Decision Tree~\cite{kdd2016}. 

In this study, using field-of-study hierarchy, we selected all the research fields within level L3 which contained at least 1,000 publications.
For each selected research field, we constructed the field's co-authorship social network over time. 
Namely, let $R$ be a selected research field, and let $y$ be a year between the time of the first and last publication in $R$.
We define the undirected co-authorship social network of $R$ at $y$ to be $G_y^{R_{co}} := <V_y^{R_{co}},E_y^{R_{co}}>$, 
where $V_y^{R_{co}}$ is a set of authors who published a paper in $R$ with a publication year before or including $y$.
In addition, each link in the dataset $e := (u,v) \in E_y^{R_{co}}$ is between two authors, 
$u \in V_y^{R_{co}}$  and $v \in V_y^{R_{co}}$, who collaborated on a 
publication in field $R$ with a publication year before or including $y$. 
Using the Microsoft Academic Graph dataset which was published for the KDD Cup 2016,
 we succeeded in constructing the social networks of 9,005 research fields over a period of 215 years. Overall, we created 770,845 co-authorship graphs. 
 
 It is important to notice that even though the co-authorship network graphs are undirected, for features 
 calculations we treated these graphs as directed graphs where each undirected link $e := (u,v) \in E_y^{R_{co}}$
  was transformed into two directed links between $u$ and $v$, and also between $v$ and $u$.

\subsubsection{The Citation Networks} 
\label{sec:citations}
Similar to the construction of the co-authorship networks described above, we utilized the Microsoft Academic Graph to 
construct the citation networks within the lowest field-of-study hierarchy category of L3. Namely, let $R$ be 
a selected research field, and let $y$ be a year between the time of the first and last publication in $R$. 
We define the directed citation network of $R$ at $y$ to be $G_y^{R_{ci} } = <V_y^{R_{ci}},E_y^{R_{ci}}>$, 
where $V_y^{R_{ci}}$ is a set of papers that were published in $R$ with a publication year before or including $y$. 
In addition, each directed link in the dataset $e := (u,v)\in E_y^{R_{ci} }$ 
is between two papers $u \in V_y^{R_{ci}}$ and $v \in V_y^{R_{ci} }$, in which paper $u$ cited paper $v$. 
Overall, we constructed the citation networks of 8,996 research fields, which include 769,793 directed graphs.

\subsubsection{The Bitcoin Transaction Network}
\label{sec:bitcoin}
Bitcoin is a cryptocurrency and a large-scale payment system, in which all the transactions are publicly accessible~\cite{ron2013quantitative}.
 In this study, we used the Bitcoin Transaction Network Dataset published in 2013 by Ivan Brugere~\cite{bitcoin}. 
 The dataset includes over 37.4 million transactions, from January 2009 to April 2013, between public-key ``addresses,'' from which we created a directed network 
with over 6.3 million vertices and 16.3 million links over a period of 222 weeks. 
Namely, we defined the Bitcoin graph at time $t$ to be $G_t^B := <V_t^B,E_t^B>$, where $V_t^B$ 
is a set of public-key addresses which perform their first transaction before time $t$, 
and $e := (u,v) \in E_t^B$ between two public-key addresses, $u \in V_t^B$  and $v \in V_t^B$, where according to the dataset a payment transaction was performed from $u$ to $v$.

\subsection{Analyzing Temporal Dynamics of Networks}
\subsubsection{Calculating Network Features}
\label{sec:features}
Throughout this study, we calculated various networks' features and analyzed how these features change over time. 
In this section, we provide formal definitions of these features. First, we define the graph of network $n$ at time $t$ to be $G_t^n := <V_t^n, E_t^n>$. 
Then, we present the following network features: 
\begin{itemize}
  \item 	\textit{Vertices number} - the number of vertices in the network at time $t$, defined as $|V_t^n|$. 
  \item \textit{Edges number} - the number of edges in the network at time \textit{t}, defined as $|E_t^n|$.
  \item 	\textit{Density} - the network's density at time $t$, defined as $D_t^n = \frac{|E_t^n|}{|V_t^n |\cdot(|V_t^n|-1)}$
  \item 	\textit{Network active time} - a network's active time (denoted as $t_{max}^n$), defined as the amount of time between the times the first and last vertices joined the network.
  \item \textit{Average clustering coefficient} - the coefficient that measures the level to which vertices in a graph tend to cluster together~\cite{saramaki2007generalizations}, defined at time $t$ 
  (denoted by $CC_t^n$) to be $G_t^n$'s average clustering coefficient.
  \item \textit{Average shortest path} - the network's average shortest path at time $t$ (denoted by $\mbox{Avg. SP}_t^n$), defined as $G_t^n$'s average shortest path. 
  \item \textit{Vertex degree} - for a vertex $v$ in network $n$, we define the vertex degree at time $t$ as 
  $d_t^n(v) = |\{u|(u,v) \in E_t^n \mbox{ or } (v,u) \in E_t^n\}|$, i.e., the number of vertices at $n$ that connect to $v$ at time 
  $t$
  \item 	\textit{K-Stars set} - using the degree definition, we define the \textit{K-Stars} set of \textit{n} at time \textit{t} 
  (denoted by $Stars_t^n(k)$) to be the set of \textit{k} vertices in \textit{n} with the highest degree at time \textit{t}.   Namely,
  \begin{eqnarray*} 
  Stars_t^n(k) &:=& \{v_1,\ldots,v_k |  d_t^n(v_i ) \geq d_t^n(v_j ) \\
   && \forall v_i \in \{v_1,\ldots,v_k\},\forall v_j \notin \{v_1,\ldots,v_k\} \\
   &&  \mbox{, and } v_i,v_j \in V_t^n\}.     
  \end{eqnarray*}
  
  \item \textit{K-Stars-Vector} - using the\textit{ K-Stars set}, we can define a network's \textit{K-Stars-Vector} over a monotonous time series 
$t_0,t_1,\ldots,t_m$ (denoted as $v_{k,n}^*$) to be the vector of size of $m$ in which each $i^{th}$ element, i.e., 
$(v_{k,n}^* )_i$, represents the number of new emerging network stars at time $t_{i+1}$. 
Namely, let there be a network $n$ which was active for time $t_{max}^n$  and let there be a monotonous time series $t_0,t_1,\ldots,t_m,\forall t_i < t_{i+1}$, 
in which $t_m \leq t_{max}^n$. We define \textit{K-Stars-Vector} over $t_i$  to be 
\[
    v_{k,n}^* = (|Stars_{t_i}^n(k)  -  \bigcup_{j=0}^{j=i-1}Stars_{t_j}^n(k) 
    |)_{i=1}^m.
\]
In creating the \textit{K-Stars-Vectors} for the networks analyzed in this study, we used a time series $t_0,t_1,\ldots,t_m$,
 in which $t_0=0$ and $t_m= t_{max}^n$, and the time difference between $t_{i}$ and $t_{i+1}$ 
  was set to one year for the co-authorship and citation networks, and typically set to 4 weeks for the subreddit networks 
  (in cases where the overall time did not divide evenly into 4-week intervals, the final interval was less than 4 weeks).
  
 \item 	\textit{K-Stars-Number} – using the \textit{K-Stars-Vector}, we can define the number of emerging stars in a network to be 
 the number of unique vertices that, at a certain time, were among the top K vertices with the highest degree in the network. 
 Namely, for a network $n$ which was active for a time $t_{max}^n$, we define the \textit{K-Stars-Number} of $n$, over time series 
 $t_0,t_1,\ldots,t_m$,  to be the sum of the \textit{K-Stars-Vector} values 
\[|v_{k,n}^* |= ||v_{k,n}^*||_1 :=\sum_{i=1}^m (v_{k,n}^* )_i.\]

\end{itemize}

\subsubsection{Vertices' Join-Time Difference}
\label{sec:methods_jrc}
Similar to Price's observation~\cite{price1965statistical} that new papers tend to be cited more than older papers,
we noticed that in all six examined networks, vertices tended to connect to vertices that joined the network at a similar time: 
(a) Reddit users tend to be more engaged with other users who joined the network at a similar time, and to be less engaged with users who became active either a long time before or after they did; 
(b) online chess players tend to play more with other players who played their first game at a similar time, and to play less frequently with those who played their first FICS game either a long time before or after they did; 
(c) in the WikiTree dataset, people tend to marry others who are about the same age, and to marry less often those with whom there is a larger age gap; 
(d) researchers tend to collaborate more with other researchers who published their first paper about the same year, and to collaborate less with researchers who published their first paper a considerable time before or after; 
(e) papers tend to cite papers more frequently that were published about the same time, and to cite older papers less frequently; and 
(f) Bitcoin transactions tend to occur more often between public-key addresses that became active about the same time, and less often between addresses that became active either a long time before or after.
 
To validate our observations, for each edge $e$ in the Reddit, chess, WikiTree, co-authorship, 
citation, and Bitcoin networks, we calculated the join-time difference between each edge's 
vertices for all the edges in our dataset. We used regression analysis 
to calculate, across all networks, the probability of a vertex $v$ connecting to a vertex $u$ as the 
function of the time difference between the join times of $u$ and $v$.

\subsubsection{Network Join-Rate-Curves}
Out of the six datasets we utilized, our study focused on the three datasets -- Reddit, co-authorship, and citation -- 
that had defined communities within the overall network, so that we could effectively analyze the evolution of their subnetwork structures.
We defined the Join-Rate-Curve of a network $n$ (denoted as $JRC_n$) to be the ratio of the number of vertices at time $t$ and the maximal number of 
vertices in the network. Namely, let $n$ be a network that was active for a time period of $t_{max}^n$; then, for $t\in[0,t_{max}^n]$, we define $JRC_n(t)\to[0,1]$ 
as:
\begin{eqnarray*}
JRC_n(t) := \frac{|V_n^t|}{|V_n^{t_{max}^n }|},  
\end{eqnarray*}

where $JRC_n(0)$ and $JRC_n(t_{max}^n )$ are defined to always be equal to 0 and 1, respectively. 

To create the JRCs for the selected networks, for each network $n$ we calculated the $JRC_n$ values using 4-week intervals
 for the subreddit networks, and using 1-year intervals for the co-authorship and citation networks of selected research fields. 
 By using these intervals, the number of samples of the JRCs for the subreddit networks ranged from 1 to 141, 
 with a median value of 51; and for both the co-authorship and citation networks ranged from 9 to 217, with a median value of 76.
 
To better understand the various types of JRCs that we created, we utilized CurveExpert software~\cite{hyams2010curveexpert} to match several selected JRCs 
 with their best-fit functions using regression analysis. To avoid overfitting, 
 we selected the best-match function with relatively low degrees of freedom. 
 Next, to verify that the selected function actually matched most
 of the JRCs, we used the python-fit package\footnote{\url{https://pypi.python.org/pypi/python-fit/1.0.0}} 
 to fit the selected best-match function on all 38,129  JRCs. 
Then, to better understand the various types of JRCs, we drew all the JRCs and ordered
  them according to the networks' vibrancies (see Section~\ref{sec:vibrancy}) 
  in descending order. Lastly, we 
  manually examined the various figure collections and scrutinized the anomalous JRCs that 
  did not fit the selected regression function in terms of $R^2$

\subsubsection{Network Vibrancy}
\label{sec:vibrancy}

 We also observed differences in topological properties among networks with different growth rates. 
 To better understand these differences, we defined the vibrancy of network $n$ to be one minus the average value of the $JRC_n$ function:
\[vibrancy(n) := 1 - \int_0^{t_{max}^n}\frac{JRC_n(t)}{t_{max}^n}.\]

The network vibrancy values range between 0 and 1, where vibrant, fast-growing networks usually have vibrancy values near 1, 
while slow-growing networks have vibrancy values near 0. 
To analyze the influence of different growth rates on the network topological properties,
we calculated the Spearman correlations among the network topological properties (presented in Section~\ref{sec:features}) and the network 
vibrancies.

\subsubsection{The Emergence  of New Network Stars}
\label{sec:methods_stars}
One of the main goals of this study was to better understand how new network
stars emerge. To achieve this, we analyzed the Spearman correlations between the frequencies at which network stars emerge
(i.e., \textit{K-Stars-Number}, for $K=1,5$) and other network properties.

Moreover, we investigated in which stage of the network’s life new stars are more likely to emerge by performing the following: 
First, for each type of network, we divided the network into two sets according to the network growth speed: fast-growing networks with vibrancies higher than $0.5$ 
($v_b>0.5$), and slow-growing networks with vibrancies lower than $0.5$ ($v_b<0.5$).
 Next, for each set, we calculated the average number of network stars which emerged in each time slice, 
 using time slices that were common for at least $w$ networks. 
 Namely, letting $N$ be a set of networks, we define the $w-maximal time$ to be the maximal time for which at least $w$ networks in the set were active:
 \begin{eqnarray*}
 t_{w,max}^N :=\mbox{ max}(\{t_{max}^n,n \in N &|& \exists n_1,n_2,\ldots,n_w \in N,  \\
 &&\forall i \in [1,w]\mbox{ } t_{max}^n \leq t_{max}^{n_i}  \}). 
 \end{eqnarray*}
Next, for a monotonous time series $t_0,t_1,\ldots,t_m$, where $t_m=t_{w,max}^n$, and for each time stamp $t_i \in \{t_1,\ldots,t_m\}$, 
we measured the average number of new stars that emerged between times $t_{i-1}$ and $t_i$ by calculating the \textit{K-Stars-Vector} 
of each network $n \in N$ over $t_0,t_1,\ldots,t_m$ and calculating the average number of 
emerging stars for each $t_i$ between $t_1$ and $t_m$. Namely, we define the following vector: 
\[TotalStarsVector_k^N := (\sum_{\{n\in N|t_i \leq t_{max}^n\}}(v_{k,n}^* )_i )_{i=1}^m, \]

where $TotalStarsVector_k^N$ is an $m$-length vector, in which each $i^{th}$  
element is the sum of the number of emerging stars at time $t_i$, across all networks in $n \in N$ 
with an active time of at least $t_i (t_i  \leq t_{max}^n)$. Then, using $TotalStarsVector_k^N$, 
we define the $AvgStarsVector_k^N$ by dividing each $i^{th}$ element in $TotalStarsVector_k^N$ 
 by the number of networks that were active for a time of at least $t_i$:
\[ AvgStarsVector_k^N :=  (\frac{TotalStarsVector_k^{N_i}}{||\{ n \in N | t_{max}^n \geq t_i \}|| })_{i=1}^m\]

Additionally, to reduce the influence of networks with frequently emerging stars on the 
$AvgStarsVector_k^N$, we also define the $NormAvgStarsVector_k^N$ as a vector with $m$ elements 
in which each $i^{th}$  element is the average normalized value of the number of emerging stars at time $t_i$ across all networks in $n \in N$:
\[ NormAvgStarsVector_k^N := (\frac{(NormTotalStarsVector_k^N )_i}{||n \in N | t_{max}^n \geq t_i\}||})   {i=1}^m,\]
where $NormTotalStarsVector_k^N$ is defined as:
\[
 NormTotalStarsVector_k^N := (\sum_{\{n \in N | t_i \leq t_max^n\}}\frac{(v_{k,n}^*)_i}{|v_{k,n}^*|})_{i=1}^m.
\]

In this study, we calculated $AvgStarsVector_k^N$ and\\
 $NormTotalStarsVector_k^N$, for $k=1,5$.

Let's take Reddit as an example. Our goal is to identify when stars tend to emerge in subreddits. To achieve this goal, we first split the subreddits into two separate groups: fast-growing subreddits ($vibrancy > 0.5$) and slow-growing subreddits ($vibrancy < 0.5$). For this example, let's focus on only the fast-growing networks (defined as $N_{f}$).  Next, we select a time series, such as $4,8,12,\ldots, 4\cdot t_m$ weeks. To avoid biasing the results from a few networks that have existed for a long time, we select a maximal time sequence ($4\cdot t_m$), which has at least $w$ active networks.

Next, we can define $TotalStarsVector_k^{N_{f}}$ as an $m$-length vector, in which each $i^{th}$ element is the sum of the number of emerging stars in all the fast-growing subreddits after $4,8,12, \ldots, 4\cdot t_m$ weeks. For this example, we will choose $k = 5$ and $t_m = 100$. Therefore, $TotalStarsVector_5^{N_{f}}$ will be a vector of size 100, in which each element in the $i^{th}$ place equals the total number of new top-5, high-degree vertices that emerged across all the fast-growing subreddits between $4\cdot (i-1)$ and $4\cdot i$ weeks since the subreddit became active. For instance, $(TotalStarsVector_5^{N_{f}})_4$ is the sum of all users who, between 12 and 16 weeks after each subreddit became active, first became one of the top-5 users in any of the fast-growing subreddits. 

Using $TotalStarsVector_5^{N_{f}}$, we have the number of stars that emerge in each time interval. However, usually more subreddits are active for shorter periods of times. Moreover, there are subreddits in which stars tend to emerge at much higher or lower rates. Therefore, we need to normalize the  $TotalStarsVector_5^{N_{f}}$, first by dividing each \textit{i}-th value by the number of networks that were active between $4\cdot (i-1)$ and $4\cdot i $ weeks (see definition of $AvgStarsVector_5^N$). Then, to reduce the influence of networks with high or low frequently emerging stars on the $AvgStarsVector_5^{N_{f}}$, we also define the $NormTotalStarsVector_5^{N_{f}}$. 

We can repeat this process on the group of slow-growing subreddits or use a similar method on other groups of networks. This methodology gives us a better understanding of when stars tend to emerge in networks' lives.

\section{Results}
\label{sec:results} 
\subsection{Vertices' Join-Time Difference }
\label{sec:results:time}

\begin{figure*}
  \centering

  \includegraphics{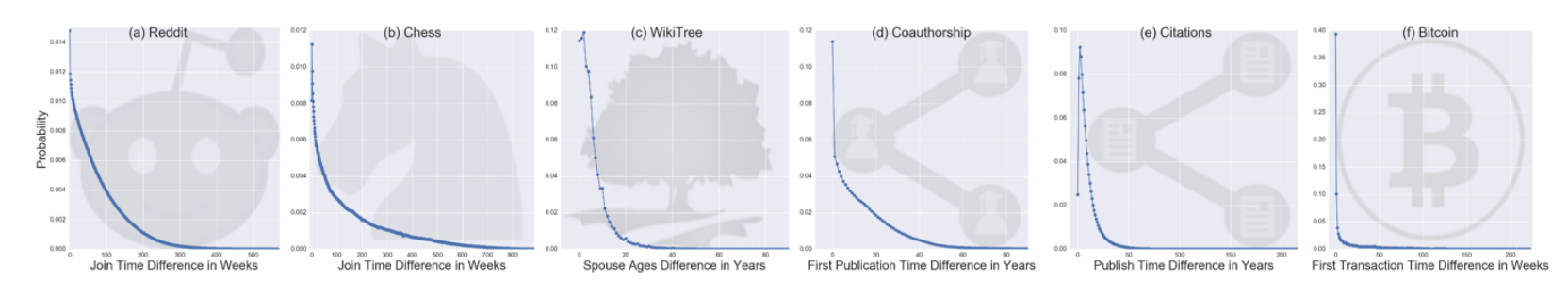}
    \caption{\textbf{The probability of two vertices connecting, as a function of the time the first joined the network.} 
    In all six real-world networks, as the join-time difference increases, the probability of vertices connecting decreases sharply,
 estimated by the following functions: (a) $0.012e^\frac{-t}{88.38}$ is the probability of a Reddit user replying to another user, where $t$ 
     is the time difference in weeks; (b) $0.0064e^{\frac{-t}{145.28}}$ is the probability of two chess players playing against each other, where $t$ 
is the time difference in weeks; (c) $0.138e^{\frac{-t}{7.01}}$ is the probability of two people getting married, where $t$ is their age difference in years;
    (d) $\frac{0.179-0.008t^{0.642}}{1.614+t^{0.642}}$ is probability of two authors coauthoring a paper, where $t$ is the time difference in years; 
    (e) $\frac{0.049+0.004t}{1-0.221t+0.04t^2}$ is the probability of one paper citing another paper, where $t$ is the years between the papers’ publications; and 
    (f) $0.391e^{\frac{-t}{0.799}}$ is the probability of Bitcoin transactions between two accounts, where $t$ is the time difference in weeks. }
  \label{fig:time_diff}
\end{figure*}

By analyzing the vertices of over 8.3 billion edges and using probability calculations and regression analysis, 
we discovered that across all networks, the probability of a vertex $v$ connecting to a vertex $u$ decreases sharply,
 typically in an exponential decline rate, as the time difference between the join times of $u$ and $v$ increases (Figure~\ref{fig:time_diff}).\footnote{The figures throughout this paper were created with high resolutions, which makes it possible to review each figure's details by zooming into the figure. }

\subsection{Network Join-Rate-Curves}
As described in Section~\ref{sec:methods_jrc}, to better understand the different 
rates in which vertices join networks, we examined and analyzed over 38,000 
JRCs.
We discovered that in most cases, the best fit was a high-degree polynomial function. 
 To avoid overfitting, we used the CurveExpert software~\cite{hyams2010curveexpert} and python-fit package to 
 find the polynomial function that was a best fit for the majority of JRCs and still had a relatively 
 low degree. We discovered that among all the subreddit networks, 18,558 (92.2\%) and 14,505 (72.06\%) of the JRCs matched quartic functions ($q(x) := a+bX+cX^2+dX^3+eX^4$) 
 with $R^2 \ge 0.95$ and $R^2  \ge 0.99$, respectively.
In addition, among all the research field co-authorship networks, 8,508 (94.49\%) 
and 5,465 (60.68\%) of JRCs matched quartic functions with $R^2  \ge 0.95$ and $R^2  \ge 0.99$, 
respectively. Furthermore, among all the research field citation networks, 8,568 (95.2\%) and 5,910 (65.7\%)
of JRCs matched a quartic function with $R^2  \ge 0.95$ and $R^2 \ge 0.99$, respectively. 

\begin{figure*}

   \centering
  \includegraphics{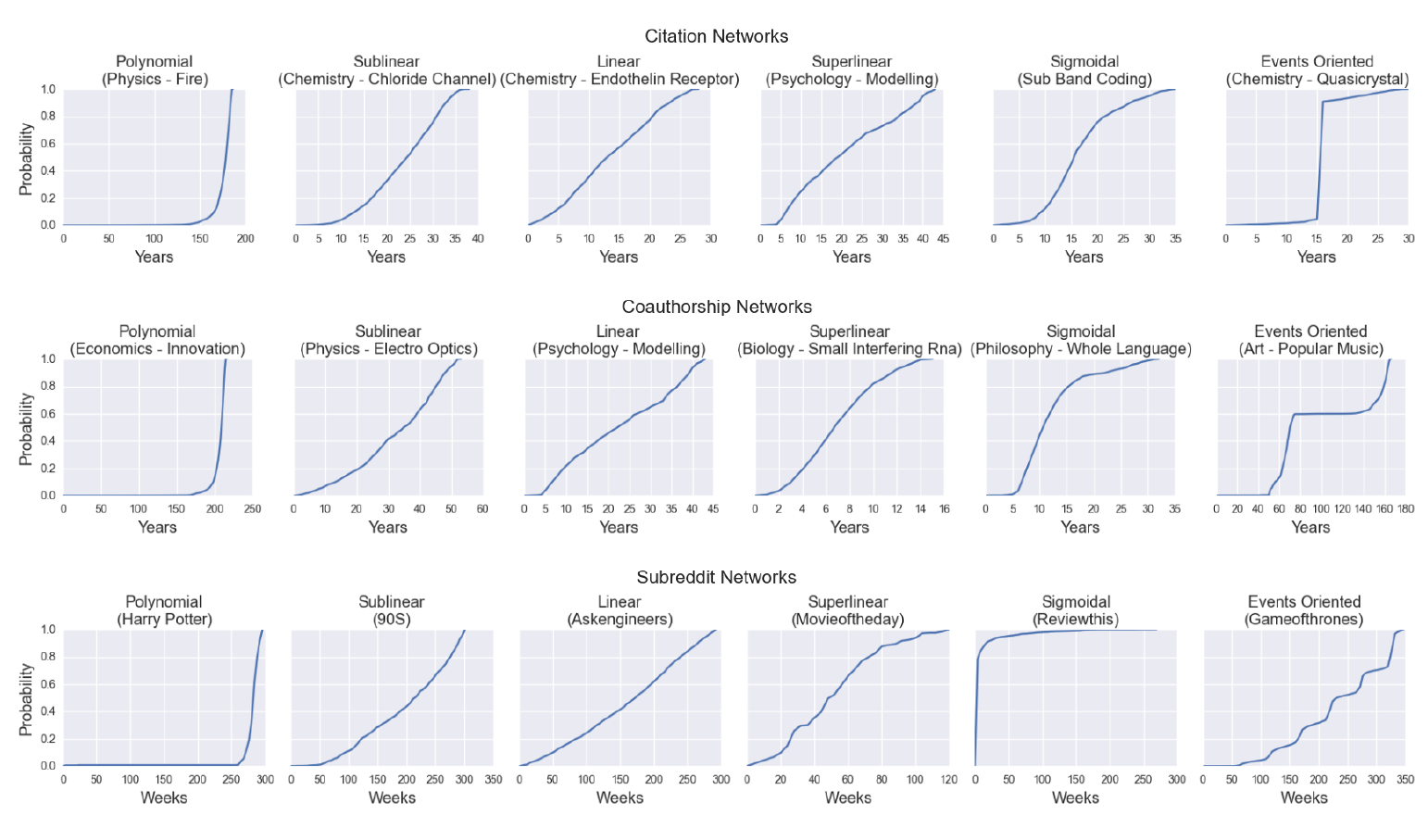}
    \caption{\textbf{Common Join-Rate-Curve patterns.} We observed five types of common JRC patterns – polynomial, sublinear, linear, superlinear, and sigmoidal. Additionally, we identified a sixth type of JRC with distinct growth patterns that are greatly affected by events, such as the quasicrystal citations network's JRC (right column) which demonstrates the field's sudden increase in popularity. These various
     growth patterns result in different network topological properties.}
      \label{fig:jrcs}
      
\end{figure*}

After observing that the vast majority of JRCs match quartic functions, our next goal was to better understand which type of quartic function the JRCs frequently match.
 We achieved this by drawing all 38,129 JRCs and ordering them according to the networks' vibrancies in descending order (see Figure Collections S1, S2, and S3). 
 Using this methodology, we observed five common JRC patterns – polynomial, sublinear, linear, superlinear, and sigmoidal (see Figure~\ref{fig:jrcs}).
  Additionally, by analyzing the JRCs that did not match quartic functions, we identified a sixth type of JRC 
  which was influenced by external events, such as the HalloweenCostume subreddit JRC that gains popularity near Halloween each year, or the JRC 
  Quasicrystal research field citation network that demonstrates an interesting growth pattern, probably due to a paradigm shift in the field.

\begin{figure*}[ht]
  \centering

  \includegraphics{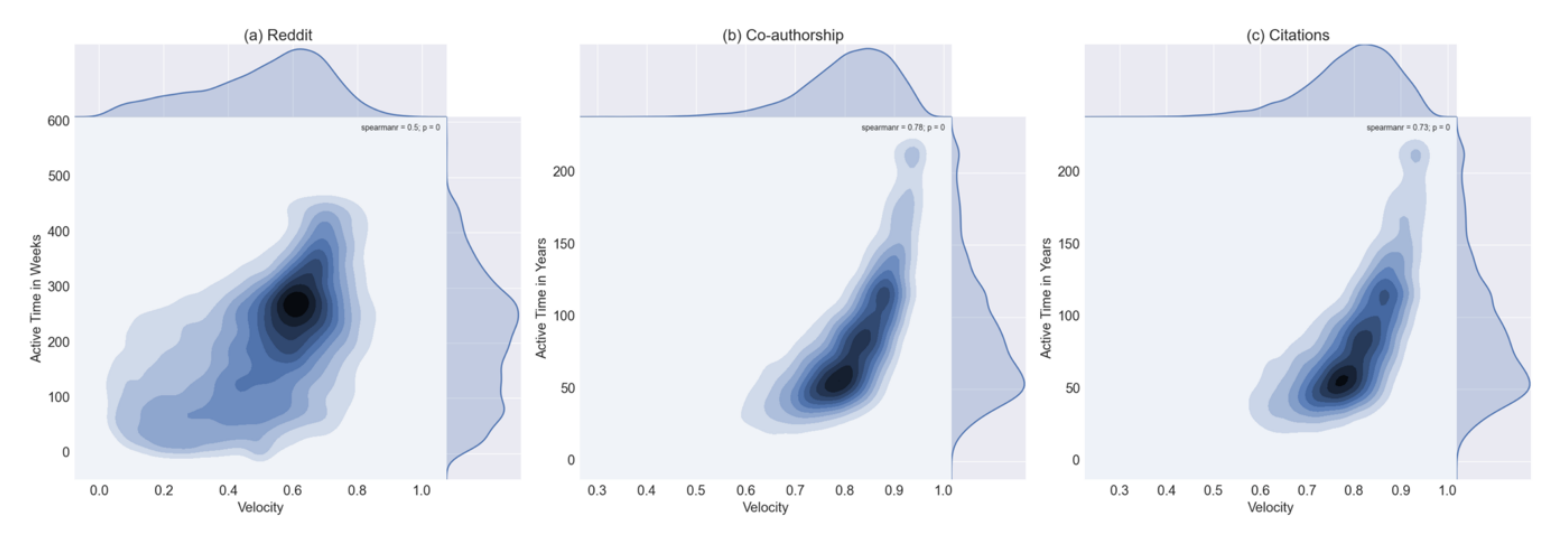}
    \caption{\textbf{Joint distributions of network vibrancy ($V_b^n$) and active time ($T_{max}^n$)}. 
    Networks with relative high vibrancy tend to be active longer. 
    Additionally, note that the subreddit networks have quite diverse vibrancy values, while the co-authorship and citation networks have mostly high vibrancy values.}
  \label{fig:joint}
  
\end{figure*}

\begin{figure*}[ht]
  \centering
  \includegraphics{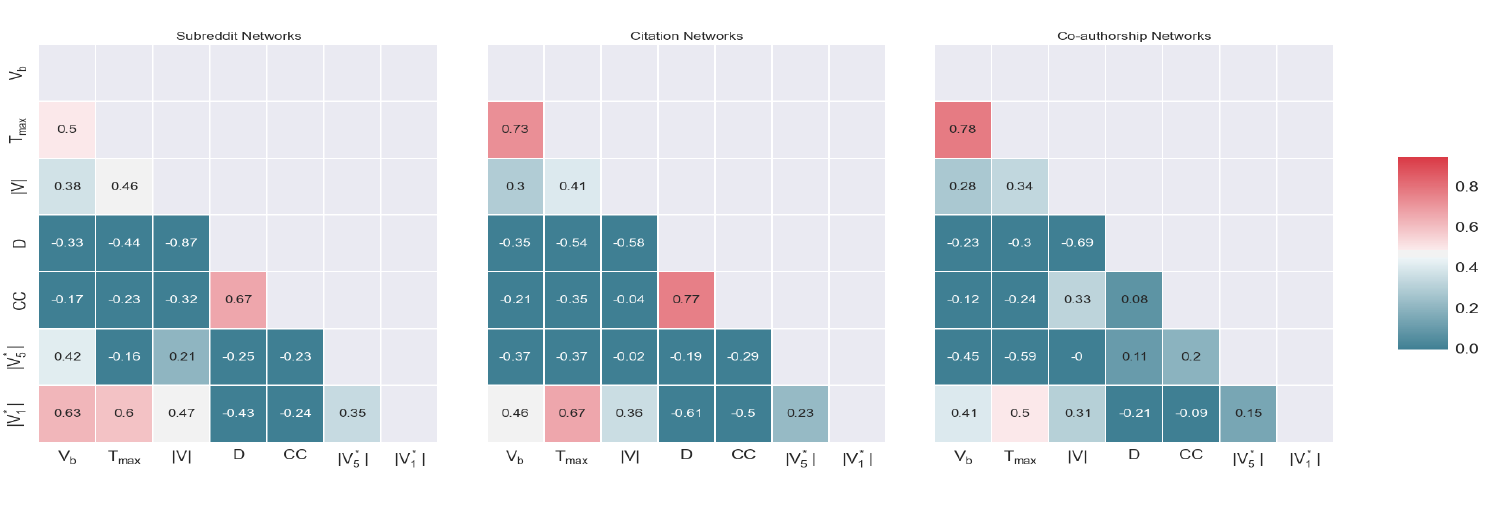}
    \caption{\textbf{Correlation matrices.}
     Correlations exist among these network features: $V_b$, vibrancy; $T$, duration the network was active; $|V|$, number of vertices; 
     $D$, density; $CC$, average clustering coefficient; $|V_5^*|$, total number of changes over time in the top-5 network stars; and 
     $|V_1^*|$, total number of changes over time in the top network star (i.e., how many times the most-linked vertex was changed).
      For these three types of networks there are high positive correlations between $V_b$ and $T$, 
      as well as medium-to-high positive correlations between $V_b$ and both $V_1^*$ and $V_5^*$, as well as between $T$ and both $V_1^*$ and $V_5^*$.}
        \label{fig:corr}   
\end{figure*}

\subsection{Network Vibrancy}
Using correlation calculations, we discovered various correlations between the vibrancies and other network characteristics (see Figures~\ref{fig:joint} and~\ref{fig:corr}). 
Primarily, we discovered the following correlations: (a) medium-to-high positive correlations $(r_s= 0.78, 0.73, 0.5)$ 
between the networks’ vibrancies and the duration in which the networks were active; (b) small-to-medium positive correlations $(r_s=0.28,0.3,0.38)$
 between the networks’ vibrancies and the
number of vertices; (c) small negative correlations $(r_s= -0.23, -0.35, -0.33)$ between the networks’ vibrancies and densities; and 
(d) small negative correlations $(r_s= -0.12, -0.21, -0.17)$ between the networks’ vibrancies and the average clustering-coefficients. 
According to these correlation results, we can discern that networks with high vibrancy tend to be active longer, have more vertices, be less dense, and cluster less than networks with low vibrancy.

\begin{figure*}[ht]
  \centering

  \includegraphics{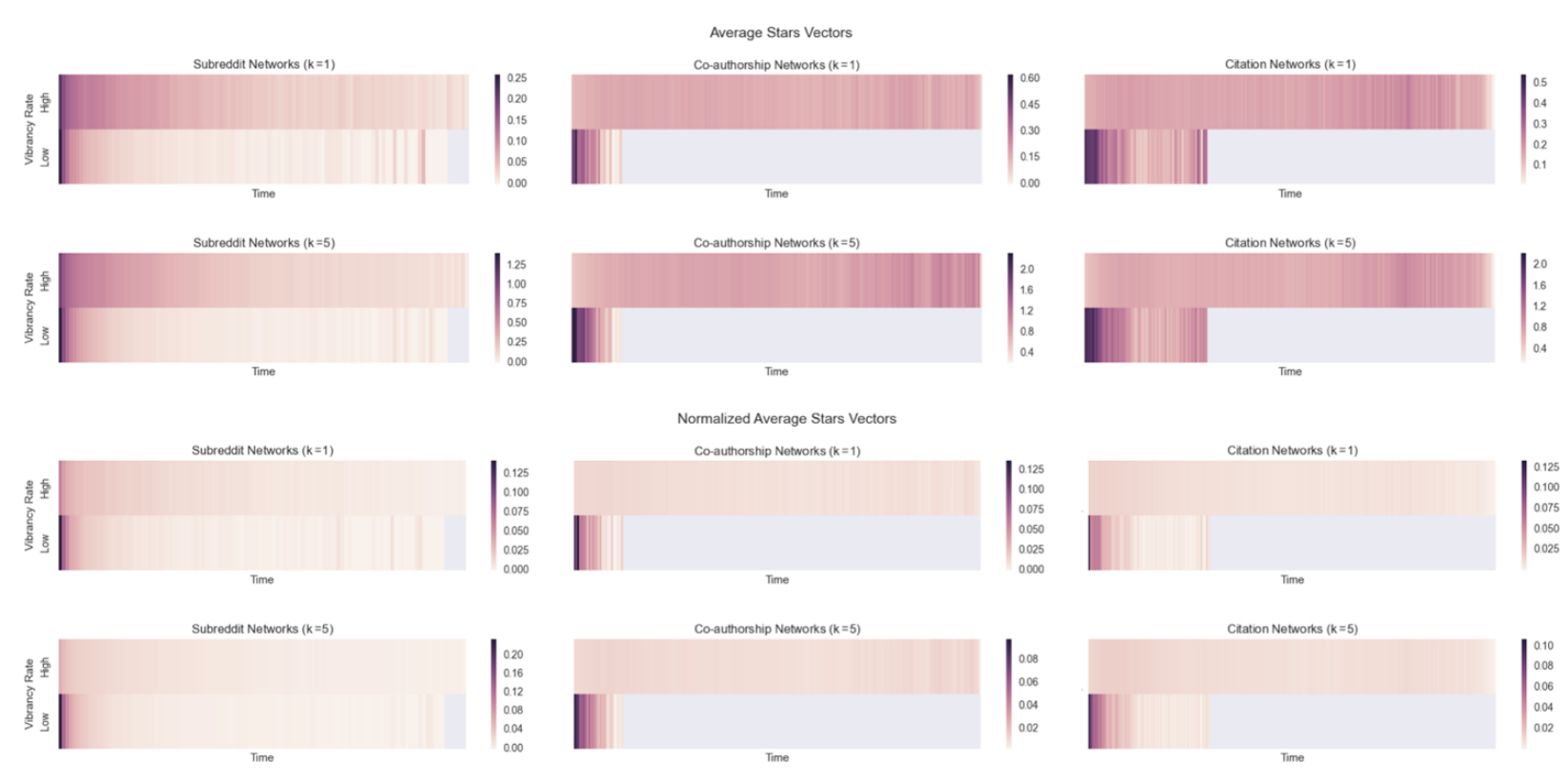}
    \caption{\textbf{New network star emergence over time.} The tendency of a new star to emerge in a network with low vibrancy is much greater 
    in the beginning than after the network matures. Additionally, for a network with high vibrancy, new stars frequently emerge at the 
    very beginning of the network’s life and tend to emerge in similar probabilities afterwards.}
  \label{fig:stars_heatmap}
  
\end{figure*}
It is worth mentioning that most of the studied co-authorship and citation networks presented relatively high vibrancy values,
 which usually indicates fast-growing networks, while the subreddit networks presented both high and low vibrancy values (see Figure~\ref{fig:joint}). 
 We believe this is a result of our research field selection process, in that we chose only successful research fields with over 1,000 published papers
  (see Sections~\ref{sec:coauthors} and~\ref{sec:citations}). 
  
\subsection{Emergence of Network Stars}
\begin{figure}[ht]
  \centering

  \includegraphics[scale=0.5]{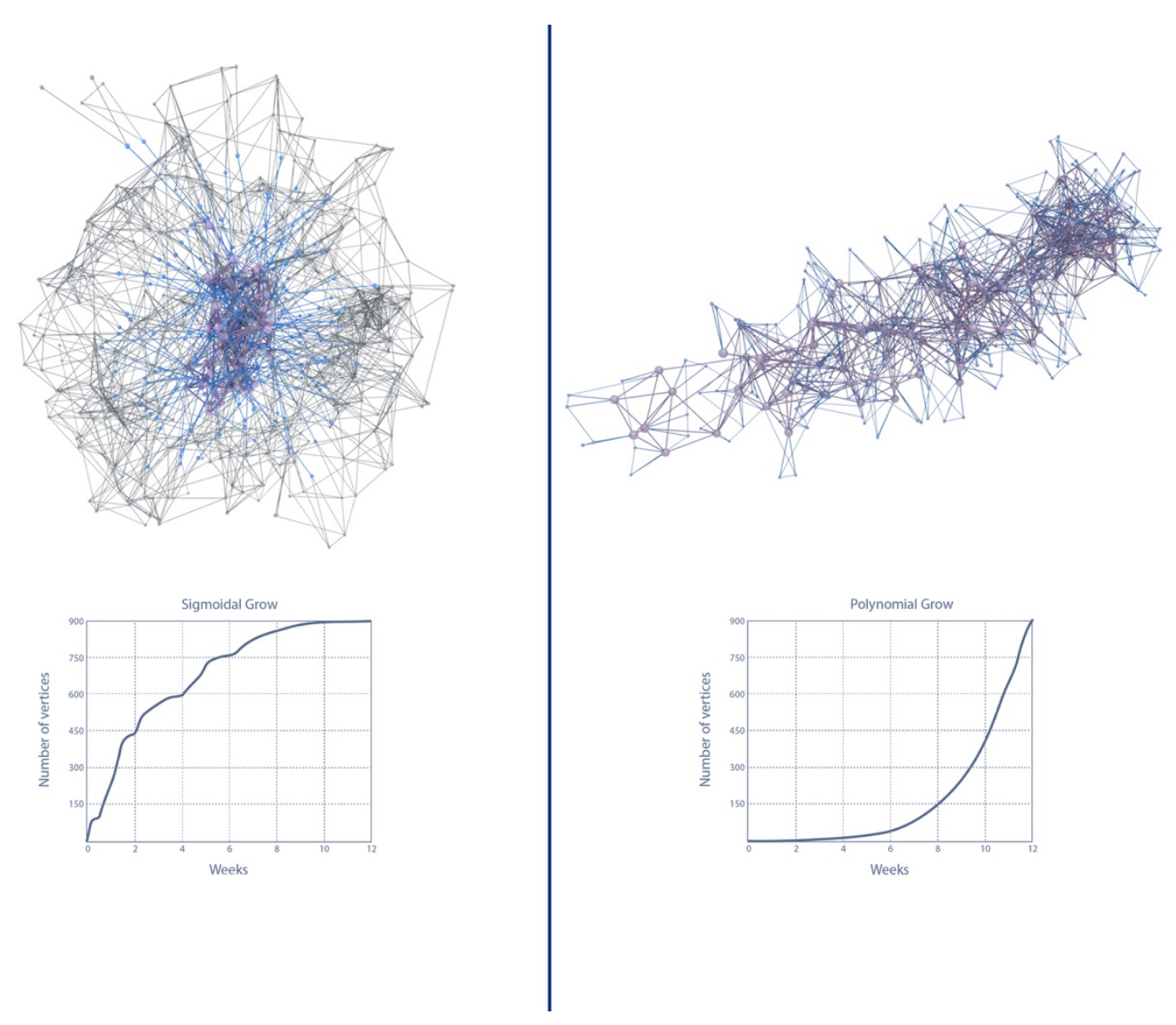}
    \caption{\textbf{Star emergence in fast- and slowgrowing networks.}
     By analyzing the network evolution process, we observed that in slow-growing networks, such as the one on the left, most stars (pink vertices) emerged a short time after the network became active and kept their place, 
     while for fast growing networks, such as the one on the right, stars emerged at any time (see Video S1). The graphs above are for illustrative purposes.}
  \label{fig:fast_slow}
  
\end{figure}

We discovered correlations between the vibrancy of networks and the changes in their most-connected vertices, i.e., their stars.
 By measuring how a list of top-5 network stars changed over time, we found medium-to-high positive correlations $(r_s=0.44, 0.44, 0.73)$ 
between the networks’ vibrancies and the total number of changes in the top-5 stars. 
Additionally, there were medium-to-high positive correlations $(r_s= 0.56, 0.71, 0.7)$ 
between the duration the networks were active and the top-5 network stars. 
Moreover, across all networks, we analyzed how the number of emerging stars changed over time (see Section~\ref{sec:methods_stars}). 
For networks with low vibrancy, most stars emerged a short time after the network became active and kept their place, while for networks with high vibrancy,
 stars emerged at any time (see Figures~\ref{fig:stars_heatmap} and~\ref{fig:fast_slow}). These results indicate that stars tend to emerge in networks that are growing rapidly.

\section{The TPA Network-Generation Model}
\label{sec:tpa}

In many complex networks the rich tend to get richer, known as the preferential-attachment process~\cite{barabasi1999emergence}.
Incorporating this tenet into the above observations, we can obtain a more complete picture of how networks evolve and how network stars emerge.
For example, consider an online social network growing at a very fast rate, i.e., with vibrancy close to 1.
As a result of the preferential-attachment process, there will quickly be several high-degree users. 
However, in line with our first and second observations, as the network continues growing rapidly,
these initial highly connected users will gain fewer connections as new-generation users will mainly connect among themselves. 
According to the preferential-attachment process, new \textit{local} generation stars will emerge among the new-generation users. 

Since the network is growing quickly, new users outnumber old users. 
Therefore, new-generation stars will eventually have more connections than old stars and will become \textit{global} stars.
This process will repeat itself as long as the network keeps growing quickly.
However, if the growth rate abruptly declines, the network will become more clustered and dense, 
resulting in fewer emerging local network stars that later become global stars.
Inspired by the above observations, we developed the TPA model, which mimics the behavior of real complex networks. 
This model generalizes the well-known Barab{\'a}si-Albert network generation model (denoted BA model)~\cite{barabasi1999emergence}
 by incorporating the role of time. Instead of adding only one vertex in each iteration, the TPA model supports the rate 
 in which vertices actually join the network, as well as the number of links each vertex establishes when it joins. 
 Moreover, the TPA model includes as input the probability that each vertex will connect to other vertices with the same or
  different join-time.   The presented TPA model produces arbitrary-sized, random scale-free networks with 
  relatively high clustering coefficients, which are sensitive to vertex arrival times and to the network’s 
  vibrancy.

\begin{figure*}[ht]
  \centering

  \includegraphics[scale=0.75]{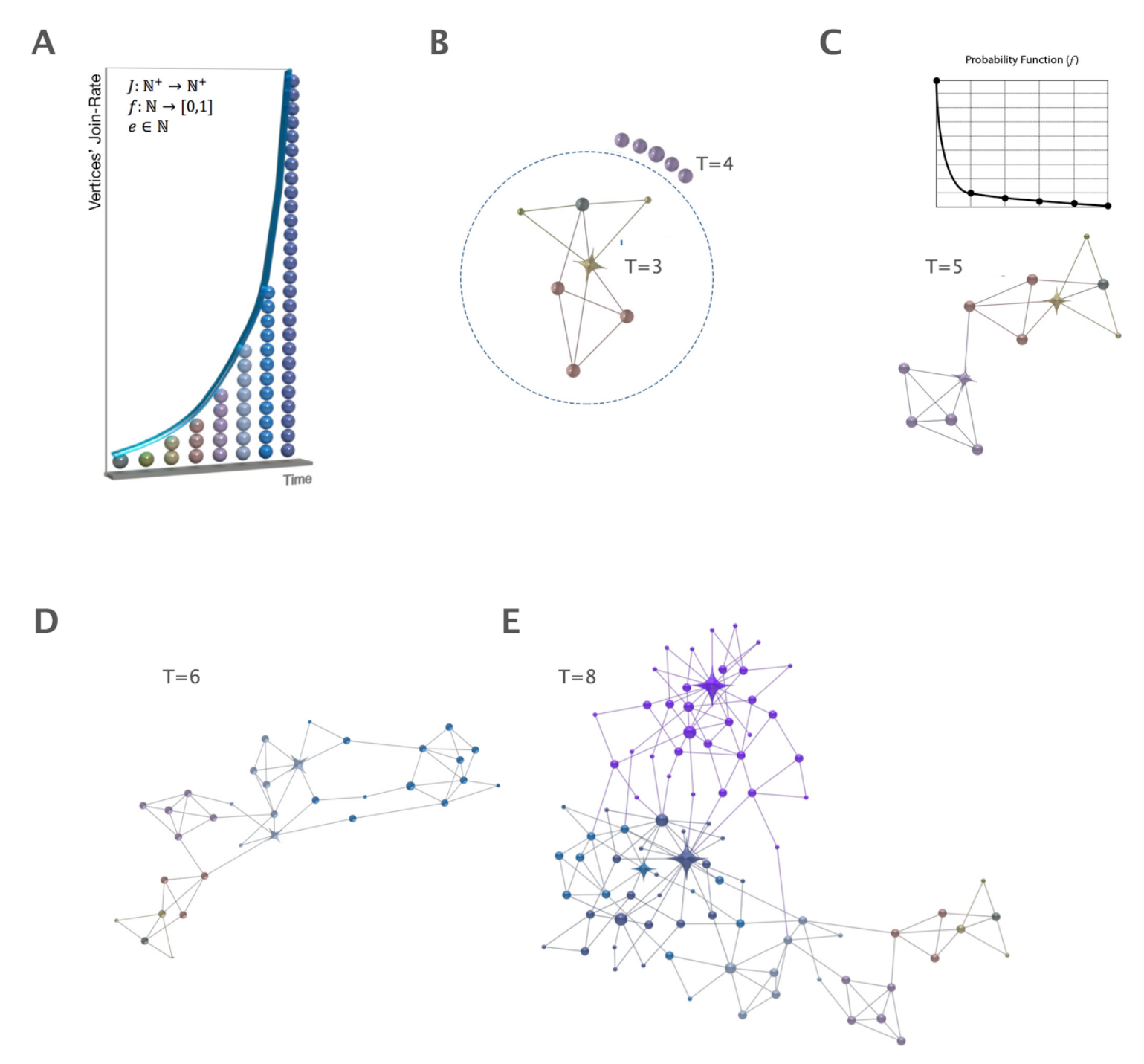}
    \caption{
\textbf{Temporal Preferential Attachment model.} The TPA model generates scale-free 
complex networks in which new stars emerge over time using the following steps: 
\textbf{(A)} There are three input parameters: \textit{j}, the vertices’ join-rate over time;  $f$, 
a monotonically decreasing function giving the probability of a vertex that arrives at time $t_i$ connecting to other vertices that arrive at time $t_j$; 
and $e$, the number of edges each vertex establishes upon joining the network. 
\textbf{(B)} The model generates a random network as, in each time iteration, a group of new vertices joins the network together (each group as a different color).
 \textbf{(C)} Each vertex $v$ establishes $e$ new links, first by selecting the time group $t_j$ to connect using the probability function $f$.
Then, a random vertex that arrived at $t_j$ is selected.
The vertex selection process is very similar to the preferential-attachment process,
i.e., a vertex is selected at random, where vertices with high degree have higher likelihood, proportional to the vertex degree, to be selected. 
Afterwards, a link is created between $v$ and the selected vertex. 
\textbf{(D)} In each iteration new groups of vertices join the network. 
\textbf{(E)} As time passes, the degree of the new joined vertices suppresses the degree of the previously joined; i.e., new network stars, marked with a star shape, are emerging.
}
  \label{fig:tpa}
  
\end{figure*}
In the following subsections, we will describe in the detail the TPA model, and the evaluate the 
properties of networks generated by TPA model's properties alongside with 
similar size networks which were generated by 
the classic BA and Small-World network models. Moreover, an implementation of the model, including code examples, can be found at the project’s website (see Section~\ref{sec:code}).  

\subsection{TPA Model Algorithm}

An overview of the TPA model algorithm is presented in Algorithm~\ref{alg:tpa} and 
Figure~\ref{fig:tpa}. The TPA model receives as input three parameters: 
first, the number of edges (denoted $m$) to attach a new vertex to existing vertices; 
second, an integers list (denoted $l$) with the number of vertices to add to the graph in each iteration; 
and third, a function (denoted $f: N \to[0,1$]) that, given a time difference value, 
returns the relative probability of an edge existing across two time groups. 
The algorithm starts by creating an empty undirected graph (line 1) and an empty time group list (line 2). 

\begin{algorithm}

  \centering
  \caption{The Temporal Preferential Attachment Model Algorithm Overview}
    \includegraphics[scale=0.8]{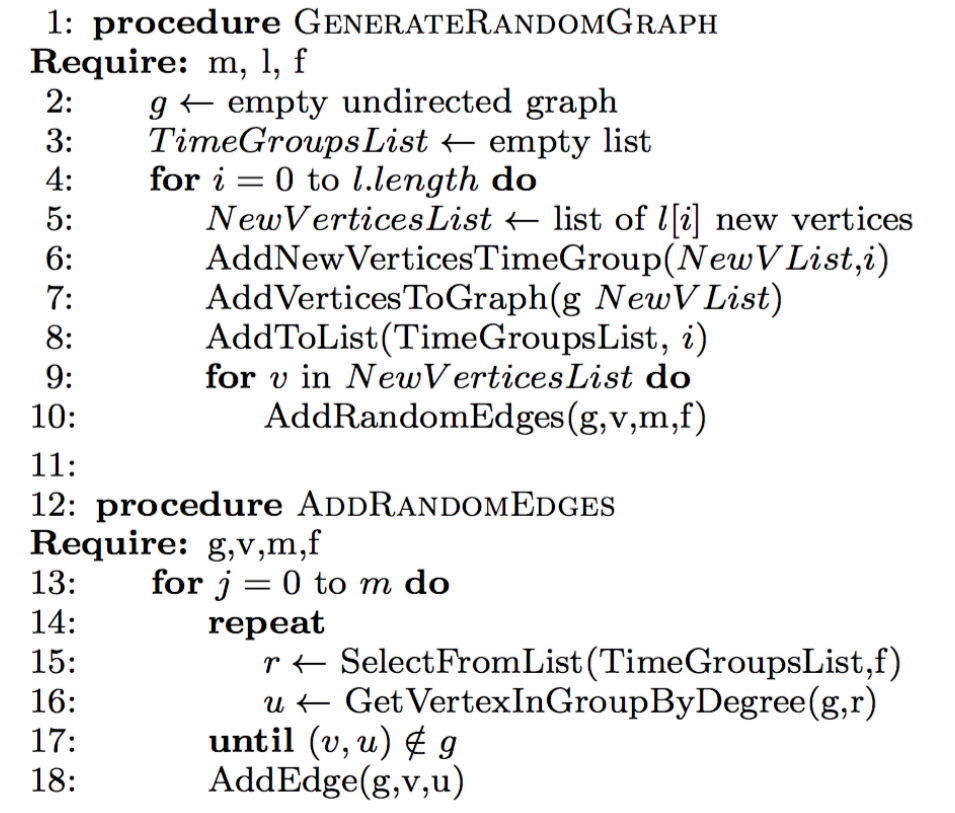}
      \label{alg:tpa}
\end{algorithm}

Then, for each positive integer $l[i]$ in $l$, the algorithm does the following:
\begin{itemize}
  \item Creates new $l[i]$ vertices with the time group set to $i$ (lines 5-6); 
  \item Adds the new vertices to the graph (line 7); 
  \item Adds $i$ to the TimeGroupsList (line 8); 
  \item Connects each new added vertex to the other $m$ vertices using the AddRandomEdges procedure (line 10).
\end{itemize}

The \textit{AddRandomEdges} procedure (lines 12-18) is the core of the model. 
The procedure receives as input five parameters: a graph ($g$), a vertex ($v$), 
the number of edges ($m$), a probability time difference function ($f$), and a list of existing time groups ($TimeGroupsList$). 
The $AddRandomEdges$ procedure connects $v$ to $m$ other vertices in the graph using the following routine:

\begin{enumerate}
  \item 	It randomly selects from $TimeGroupsList$ a time group (denoted $r$) where the probability of selecting
   each time group is given by $f$ (line 1), where given $t_1,t_2,\ldots,t_n$ time groups,
   the actual probability of an edge being created between two time groups with a time difference of $d\le n$ is equal to $\frac{f(d)}{\sum_1^n f(t_i)}$.

  \item Similar to the BA model, the procedure selects one vertex ($u$)
   among all the vertices that are in the selected time group $r$,
    where vertices with higher degree have higher likelihood of being selected (line 16). 
    In case the edge $(u,v)$ already exists in the graph, then the selection process of $u$ is repeated until a new $u$ in the graph is created.\footnote{
    In the Python implementation of the TPA model (see Section~\ref{sec:code}), we limited the number of repeats to prevent cases where it is impossible to add new edges to $v$.}
    
\end{enumerate}

To illustrate our TPA model algorithm, we can create a random graph using the following input parameters:
$m=3$, $l=(100,200,400)$, and $f(t)=2^{-1-t}$. We start running the model with an empty graph.
In the first iteration, we add $100$ ($l[0]$) new vertices to the graph, and each new vertex has a time group value of $0$. 
In this iteration there are not any other time groups. Therefore, the $100$ new vertices will create only $300$ $(100\cdot 3)$ edges among themselves in the 
following manner: 
each vertex will select $3$ other vertices in the group, and, similar to the BA model, 
vertices with higher degree will have higher probability of being selected, i.e., 
the richer vertices will have a higher probability of becoming richer.

\begin{figure*}
  \centering

  \includegraphics[scale=0.9]{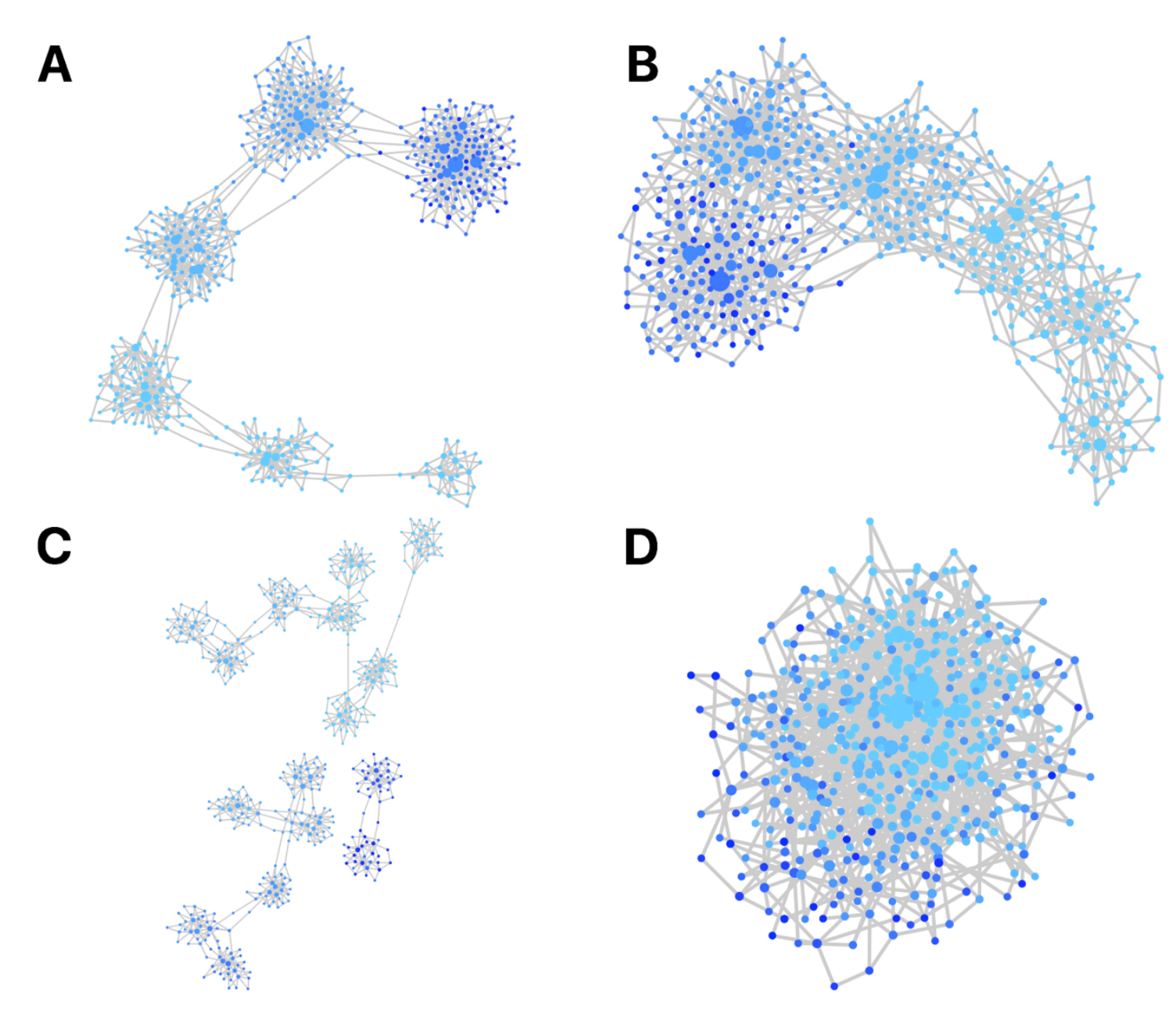}
    \caption{\textbf{Networks with various topological structures created by the TPA model.} 
    Networks \textbf{A} and \textbf{B} were constructed using a fast growth rate and setting the time difference functions to create 98\% and 86\% 
    of links in the same time group, respectively. Network \textbf{C} was created using constant growth by adding 30 vertices to the network in each iteration
     and setting the time difference functions to create 95\% of the links in the same time group. 
     Network \textbf{D} was created using a sigmoidal-like growth rate and setting the time difference functions to create 65\% 
     of the links in the same time group. In all four graphs the number of edges ($m$) was set to $2$. 
     Additionally, the color of the vertices in the graph represents the time in which the vertices were added to the networks; 
     light blue vertices were added earlier than dark blue vertices. 
     Also, the size of each vertex is proportional to degree of the vertex, i.e., larger vertices have higher degrees.}
  \label{fig:tpa_example}
  \end{figure*} 

In the second iteration, the model will insert $200$ $(l[1])$ new vertices, 
which will form $600$ $(200\cdot 3)$ new edges. However, this time we have two time groups: 
(a) a time group of $1$ (with time difference $0$), which contains all the $200$ new vertices,
and according to the time difference probability function, the probability of each new vertex establishing
a connection to this group is $f(0)=2^{-1-0}=0.5$; and (b) a time group of $0$ (with time difference of $1$), 
which contains the previous $100$ vertices, with a probability of $f(0)=2^{-1-1}= 0.25$ of 
connecting to vertices in this time group. 
According to these parameters, we can observe that the probability ratio of the two time groups is $2$ to $1$. 
Therefore, we can use this ratio to estimate that out of the $600$ edges of the second iteration, about $400$ edges will be formed among the vertices of time group $1$, 
and about $200$ edges will be formed among the vertices of time group $1$ and time group $0$, 
where each edge has a higher probability of connecting vertices with higher degree.
A detailed implemented TPA model in Python can be found in the paper's website

Lastly, in the third iteration, our model will insert an additional $400$ $(l[2])$ 
new vertices to the graph with a time group value of $2$. These vertices will formulate $1,200$ $(400\cdot 3)$ edges, 
of which about $686$ will be among the vertices of time group $2$; about $343$ edges will be among the vertices of time 
groups $2$ and $1$ (time difference of $1$); and about $171$ edges will be among the vertices of time groups $2$ and $0$ (time difference of $2$). 
Overall, the TPA model will have constructed a graph with $700$ vertices and $2,100$ edges.

\subsection{TPA Model Evaluation}
To empirically evaluate the TPA model, we created various random networks using various input parameters: 
The vertices number was set to three different sizes: $700, 6,200$, and $12,350$.
The edge number, parameter $m$, was set to $3$, creating networks with about $2,100, 18,600$, and $37,050$ edges.
We used linear, polynomial, and sigmoidal vertex growth rates. For the linear growth rate, we added $10$ 
new vertices in each iteration. For the polynomial growth rate, we used the sequence of $5,20,45,...,5x^2$, 
with a maximal $x$ value of $8, 16$, and $20$ for creating networks with $2,100, 18,600$, and $37,050$ edges, respectively. 
For the sigmoidal growth rate, we used the same growth sequence as used in polynomial growth, only in reverse order.
We used $f(t)=2^{-1-t}$ and $f(t)=0.8\cdot 0.2^t$ functions as time difference functions ($f$), where $f(t)=0.8\cdot0.2^t$
 will create considerably more edges among all the vertices in the same time group than $f(t)=2^{-1-t}$.
 
 \begin{table*}
  \centering
    \caption{Random Networks' Topological Properties.}
  \includegraphics[scale=0.78]{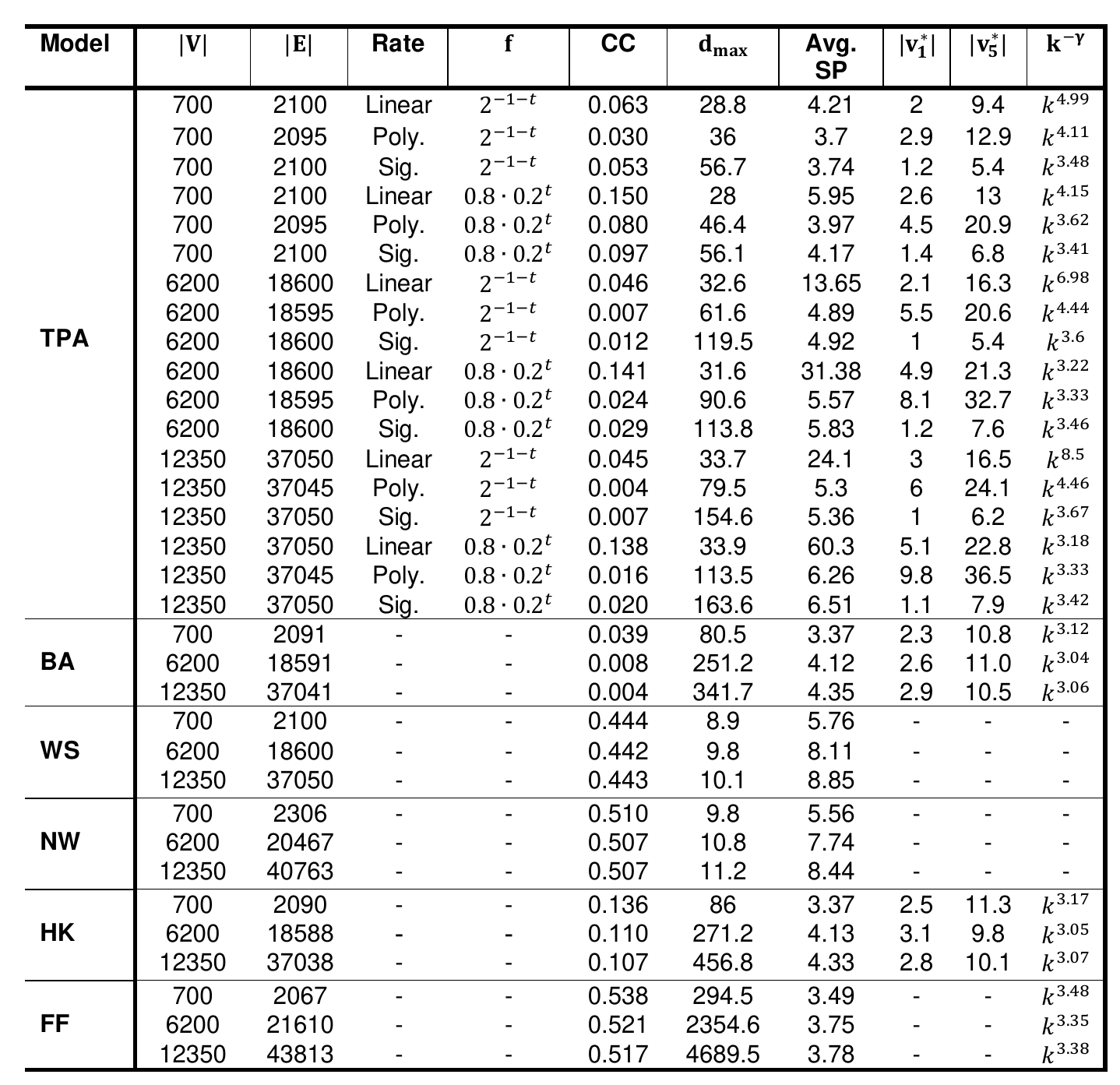}

  \label{tab:tpa}
\end{table*}

Overall, we assembled 18 different parameter settings for generating random networks. 
For each parameter setting, we utilized the TPA model to create 18 random networks.
 Subsequently, for each network, we calculated the network's average clustering coefficient, the maximal degree of vertex in the network, the network's average shortest path value, 
 the \textit{K-Stars-Number} for $k=1,5$ (denoted $v_1^*$ and $v_5^*$), and the power-law function ($k^{-\gamma}$) that matched 
the degree distribution of the network. 
To reduce variance of the calculated features, we repeated the network construction process and feature calculations $10$ times for each parameter setting and calculated the average value of each feature. 
The results of these calculations are presented in Table~\ref{tab:tpa}. 
Furthermore, for comparing the TPA model with other models, we used the BA model~\cite{barabasi1999emergence}, the Watts-Strogatz model (denoted WS) ~\cite{strogatz2001exploring}, the Newman-Watts model (denoted NW)~\cite{newman1999renormalization}, the Holme and Kim model (denoted HK)~\cite{holme2002growing}, and the Forest Fire model (denoted FF)~\cite{leskovec2007graph}.
To generate random networks of similar sizes, where the $p$ parameters in the WS and NS model was set to 0.1, in the HK model the  probability of adding a triangle after adding a random edge was set to 0.2, and in the FF model the forward probability was set to 0.65 (see Table~\ref{tab:tpa}).\footnote{For the BA, WS, NW, and HK models, we utilized the graph generation code from the Networkx package (see \url{https://networkx.github.io/documentation/latest/reference/generators.html}. We used the IGraph package implementation for the FF model (see \url{http://igraph.org/c/doc/igraph-Generators.html}).}

\section{Discussion}
\label{sec:diss}
By analyzing the results presented in Sections~\ref{sec:results} and~\ref{sec:tpa}, the following can be noted:

First, as can be observed in Section~\ref{sec:related}, the field of complex networks is
 flourishing, with an ever-growing body of work and an increasing number of random network generation models. The massive corpora of networks created and released due to this study can greatly contribute to a better understanding of complex dynamic networks, both by identifying which existing models best reflect real-world networks and by helping create models which more accurately mimic real-world network behavior.

Second, by examining the JRCs of over 38,000 networks, we discovered six
main common network growth patterns. We showed that there are notable differences between the 
structural properties of polynomial-growing networks and of sigmoidal-growing 
networks. 

Third, as observed in our data, the time and rate in which vertices join a network have a crucial effect on the network's structure and dynamics.
For example, as shown in Figure~\ref{fig:corr}, fast-growing networks with high vibrancies 
and slow-growing networks with low vibrancies tend to present different topological features. 
This observation is also supported by networks created with the TPA model, where different time and rate parameters produce networks with different topological properties.
Figure~\ref{fig:tpa_example} provides generated examples showing the variety of networks with different topologies that can be created using different time and rate parameters. 
In addition, we can observe that fast-growing networks with high vibrancies tend to be active longer than slow-growing networks with low vibrancies (Figure~\ref{fig:joint}).
Therefore, the time and rate in which networks evolve are two key factors that must be included in understanding complex networks.

Fourth, network stars emerge differently in fast- and slow-growing networks (see Figures~\ref{fig:stars_heatmap} and~\ref{fig:fast_slow}). 
In slow-growing networks, most stars emerge a short time after the network becomes active and keep their place, while in fast-growing networks, stars emerge at any time. Furthermore, based on the TPA model and our other observations, we can predict the chances of a new star surpassing an old star.

Fifth, networks with low vibrancies typically have higher average clustering coefficient ($CC$) 
values than networks with high vibrancies (see Figure~\ref{fig:corr}). 
This is confirmed by the networks generated with the TPA model, 
in which networks created by sigmoidal growth usually presented higher $CC$ values than same-size networks created by polynomial growth (see Table~\ref{tab:tpa}).

Sixth, it is important to keep in mind that community networks can affect each other within a larger network~\cite{hessel2016science}. For example, sudden growth in one research community can result in slowing growth in another research community. Even unconnected networks can influence each other. For example, even though Facebook, Twitter, and WhatsApp are different social platforms, they can considerably influence the network properties of each other.
In future research, we plan to study the connections among various communities' growth patterns.

Seventh, unlike many other network-generation models, the TPA model is sensitive to the rate and the time in which vertices join the network.
Furthermore, similar to the BA model, the TPA model also takes into account the degree of each vertex, where high-degree vertices have a higher likelihood of being connected to new vertices.
Additionally, the TPA model generates scale-free networks with similar degree distribution to networks created by the BA model,
but with much higher CC values than the BA model (see Table~\ref{tab:tpa}). The CC values obtained by the TPA model's networks indicate that vertices tend to cluster more than in networks created by the BA model, and the TPA values are more similar to the values presented in networks created by the HK model. Additionally, the TPA model can create networks with relatively small average shortest path values.\footnote{According to Table~\ref{tab:tpa}, the TPA model is able to generate networks with AVG. SP values that are smaller than AVG. SP values of similar sized networks generated by WS and NW models, and slightly higher than AVG. SP values of networks generated by BA, HK, and FF models. }

In addition, we can notice that the power-law function ($k^{-\gamma}$) of networks created by the TPA model often has similar values to similar-size networks created by other models, with $\gamma \in [3,4]$ values. In addition, it can be observed that networks created by the TPA model usually have maximal degree ($d_{max}$) which is considerably less than in networks created by the BA, HK, and FF models. This observation indicates that links in the networks created by the TPA model are less globally governed by the ``rich-get-richer'' rule than other models. Therefore, unlike in many other models, in networks generated by the TPA model new stars can emerge in any time. Moreover, the TPA model can generate random networks with diverse topologies (Figure~\ref{fig:tpa_example}). 
Additionally, while most vertices with high-degree in the BA model likely joined the network in the first iterations, 
the TPA model's vertices joined the network in later iterations. This more accurately mimics a real-world network's evolution 
process and provides insight on how a newly added vertex can suddenly become popular, such as when a post becomes viral in social networks.
While the TPA model  explains the rate and the time vertices join the network, as well as how the preferential attachment process 
influences network topology, there are other factors, such as vertex and edge properties, that may also considerably influence the topology.
We believe that with
future releases of additional real-world temporal complex network datasets, we will be able to utilize
additional data and refine the TPA model in simulating real world
networks.

Eighth, the TPA model and the study's observations can provide guidelines on where to look for the next rising network stars. In slow-growing networks, we can assume that the stars of the past will very likely continue as the stars of the future. In fast-growing networks, we can predict that new stars will rise with every new generation. According to our observations, most of the future stars will be linked with other vertices that joined the network in a similar time generation. Therefore, a practical approach to identifying future stars is to scout for new ``local'' stars, i.e., vertices which have joined the network recently and also are connected in a high degree to other vertices that joined the network at a similar time. Another conclusion we can derive from our study is that in fast-growing networks, change is indeed inevitable, and the network stars of the past will likely fall and be replaced by new stars.

Ninth, according to the TPA model, we can predict that in high-vibrancy networks, if only one new generation decides not to join the network, it may have a destructive effect on the growth of the network due to the tendency of new edges to emerge mainly among  vertices that join the network at similar times. Moreover, due the many edges among close generations, the effect of one generation leaving the network can hugely affect future generations that may also decide to leave the network. These type of changes can help explain how fast-growing networks become slow-growing networks.

Lastly, it is important to emphasis the significance of analyzing datasets that are large in scale when uncovering the evolution process of complex networks. Without large-scale analysis, it would be very challenging to identify the existence of the various JRCs, especially the existence of ``events-oriented'' JRCs. Moreover, it would be difficult to understand the effect of different JRCs on the network structure and on the emergence of network stars. It is essential to apply cutting-edge data science tools and to use large datasets  to gain important insights on the evolution process of complex networks.

\section{Conclusions}
\label{sec:con}
The field of data science has undergone many recent advances, and new algorithms, infrastructures, and techniques for data mining, data storage, data prediction, and data visualization have emerged~\cite{low2012distributed,bostock2011d3,zaharia2010spark,armbrust2010view}.
These tools make it feasible to gain new insights from vast quantities of data. 
In this study, we utilize data science tools to construct the largest publicly available network evolution corpora to date, in order to 
perform the first precise wide-scale analysis of the evolution of networks with various scales. Our study uses real data from actual networks.

We utilized the corpora to deeply examine the evolution process of networks and to understand how popularity shifts from one vertex to another over time. 
From our analysis, three key observations emerged: First, links are more likely to be created among vertices that join a network at a similar time. 
Second, the rate in which new vertices join a network is a central factor in molding a network's topology. 
Third, the emergence of network stars, i.e., high-degree vertices, is correlated with fast-growing networks. 
Based on these observations, we have developed a simple, random network generation model. 
Our Temporal Preferential Attachment (TPA) model more closely represents real-world data in fast-growing networks than previous models, many of which used a relatively small amount of data or only partial real data.

Moreover, the large corpus of networks created and released due to this study can greatly contribute to a better understanding of complex networks in general. 
We endorse the words of Albert-L{\'a}szl{\'o} Barab{\'a}si: ``If data of similar detail capturing the dynamics of processes taking place on networks were to emerge in the coming years, our imagination will be the only limitation to 
progress.''~\cite{barabasi2009scale}
Much progress is being made in the field of complex networks, and our research emphasizes the value of using vast quantities of real data to create models that accurately represent the world around us. We must stay true to the real world to keep progressing in the right direction.

\section{Data and Code Availability}
\label{sec:code}
\begin{figure*}
  \centering

  \includegraphics[scale=0.9]{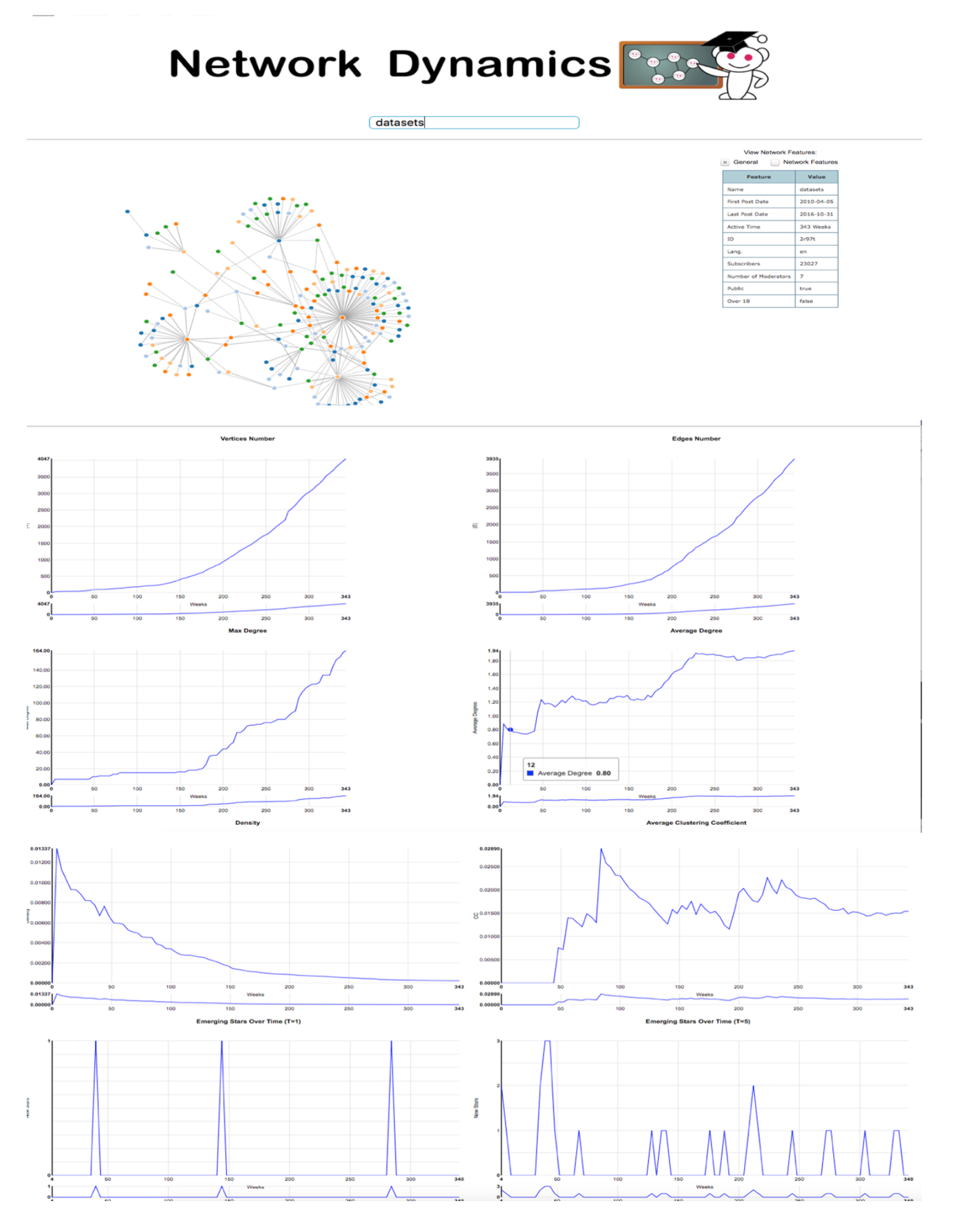}
    \caption{We have developed an interactive \href{https://dynamics.cs.washington.edu}{\underline{website}} that makes it possible to view and interact directly with the study’s data.}
  \label{fig:website}
  \end{figure*}

One of the main goals of this study was to create the largest complex network evolution public dataset. 
Therefore, the Reddit, FICS Games, WikiTree, Microsoft Academic Graph, and Bitcoin Transaction datasets 
used to create the networks and graphs in this study are all open and public. The social network datasets 
and a considerable part of the study's code, including implementation of the TPA model and code tutorials,
 are available at the project's \href{https://dynamics.cs.washington.edu}{\underline{website}} which also gives researchers the ability to interactively
  explore and better understand the networks in this study's dataset (see Figure~\ref{fig:website}).

\phantomsection
\section*{Acknowledgments} 

\addcontentsline{toc}{section}{Acknowledgments} 
First and foremost, we would like to thank Jason Michael Baumgartner, Chris Whitten, Ivan Brugere,
FICS Games Database team, and the Microsoft Academic Graph team for making 
their datasets available online. Additionally, we thank the AWS Cloud Credits for Research.
We also thank the Washington Research Foundation Fund for Innovation in Data-Intensive Discovery, and the
Moore/Sloan Data Science Environments Project at the University of Washington for supporting this
study. Datasets, software implementations, code tutorials, and an interactive web interface for 
investigating the studied networks are available at \url{https://dynamics.cs.washington.edu/
}

We also wish to thank Carol Teegarden for editing and proofreading this article to completion. Additionally, we thank Danilo Gutierrez for creating the project’s video and Aaron Romm for contributing his voice to the video. We also thank Stephen Spencer for his IT expertise. Lastly, we wish to thank Dima Kagan and Hila Fire for their assistance during this research.

\phantomsection
\bibliographystyle{unsrt}
\bibliography{stars}


\appendix
\section{Supplementary Multimedia Links}
\subsection{External Datasets}
The following project’s datasets are available at the project’s website:
\\
\textbf{Dataset S1.} The Reddit networks’ evolution dataset. This dataset contains the evolution over time of 20,128 subreddits and their corresponding 1,023,995 graphs (about 478 GB of compressed data).\\
\textbf{Dataset S2.} The Free Internet Chess Server network's evolution dataset (about 6.4 GB of compressed data).\\
\textbf{Dataset S3.} The co-authorship networks’ evolution dataset. This dataset contains the co-authorship 9,005 networks of research fields and their corresponding 770,854 graphs (about 419 GB of compressed data).\\
\textbf{Dataset S4.} The citations networks’ evolution dataset. This dataset contains the citation networks of 8,996 research fields and their corresponding 769,793 graphs (about 29 GB of compressed data).\\
\textbf{Dataset S5.} The Bitcoin Transaction network's evolution dataset (about 0.6 GB of compressed data).\\
\textbf{Dataset S6.} The Reddit networks’ final graphs dataset. This dataset contains the final graph instance of 20,128 subreddits in October 2016 (about 25 GB of compressed data).\\
\textbf{Dataset S7.} The Join-Rate-Curves dataset. This dataset contains 38,129 times-series of the co-authorship, citation, and subreddit JRCs analyzed in this study.

\subsection{Code Tutorials}
The following code tutorial are available at the project’s website:\\
\textbf{Code Tutorial S1.} Analyzing the social networks of over 2.7 billion Reddit comments. A Jupyter Notebook code tutorial explains and demonstrates how we analyzed the Reddit dataset.\\
\textbf{Code Tutorial S2.} The TPA model code. A Jupyter Notebook code tutorial provides explanations of how to create random complex networks using the TPA model.

\subsection{Figure Collections}
The following figure collections are available at the project’s website:\\
\textbf{Figure Collection S1.} The citation networks’ Join-Rate-Curves. This dataset consists of 8,996 citation network JRCs ordered by the networks’ vibrancies in descending order.\\
\textbf{Figure Collection S2.} The co-authorship networks’ Join-Rate-Curves. This dataset consists of 9,005 co-authorship network JRCs ordered by the networks’ vibrancies in descending order.\\
\textbf{Figure Collection S3.} The subreddit networks’ Join-Rate-Curves. This dataset consists of 20,128 subreddit network JRCs ordered by the networks’ vibrancies in descending order.

\subsection{Supplementary Video}
\textbf{Video S1.} The Rise and Fall of Network Stars Video. This 4:10-minute video provides an overview of both the main research results and of the TPA model.

\end{document}